\documentclass[
10pt,
aps,
prl,
amsmath,
amssymb,
twocolumn,
superscriptaddress]{revtex4-2}

\usepackage{graphicx}
\usepackage{dcolumn}
\usepackage{bm}
\usepackage{xcolor} 
\usepackage{placeins}
\usepackage[linkcolor = blue, citecolor = red, urlcolor = blue, colorlinks = true]{hyperref}

\begin{document}


\title{Activity drives self-assembly of passive soft inclusions in active nematics}
\author{Ahmet Umut Akduman}
\thanks{These authors contributed equally.}
 \affiliation{Physics Department, College of Sciences, Ko{\c c} University, Rumelifeneri Yolu 34450 Sar\i{}yer, Istanbul,  T{\" u}rkiye}
 
\author{Yusuf Sar\i{}yar}
\thanks{These authors contributed equally.}
 \affiliation{Physics Department, College of Sciences, Ko{\c c} University, Rumelifeneri Yolu 34450 Sar\i{}yer, Istanbul,  T{\" u}rkiye}

\author{Giuseppe Negro}
\email{giuseppe.negro@ed.ac.uk}
 \affiliation{SUPA, School of Physics and Astronomy, University of Edinburgh, Peter Guthrie Tait Road, Edinburgh, EH9 3FD, UK}
 
\author{Livio Nicola Carenza}
\email{lcarenza@ku.edu.tr}
 \affiliation{Physics Department, College of Sciences, Ko{\c c} University, Rumelifeneri Yolu 34450 Sar\i{}yer, Istanbul,  T{\" u}rkiye}

\begin{abstract}

Active nematics are out-of-equilibrium systems in which energy injection at the microscale drives emergent collective behaviors, from spontaneous flows to active turbulence. While the dynamics of these systems have been extensively studied, their potential for controlling the organization of embedded soft particles remains largely unexplored. Here, we investigate how passive droplets suspended in an active nematic fluid self-organize under varying activity levels and packing fractions. Through numerical simulations, we uncover a rich phase diagram featuring dynamic clustering, activity-induced gelation, and a novel \emph{inverse} motility-induced phase separation regime where activity stabilizes dense droplet assemblies. Crucially, we demonstrate that temporal modulation of activity enables precise control over structural morphological transitions. Our results suggest new routes to design adaptive smart materials with tunable microstructure and dynamics, bridging active nematics with applications in programmable colloidal assembly and bio-inspired material design.
\end{abstract}

\maketitle

Active nematics are a class of non-equilibrium systems that consume energy at the microscopic scale, driving them far from equilibrium and enabling the emergence of spontaneous flows and complex macroscopic behaviors~\cite{marchetti,sanchez2012,review_autonomous_materials,doostmohammadi2018}. At low activity levels, energy injection leads to local deformations of the active nematics, resulting in ordered flow patterns~\cite{Voituriez2006,giomi2011,doostmohammadi2018}. However, as the rate of energy injection increases,  deformations strengthen, accompanied by a corresponding rise in the magnitude of active flows. This drives the system into a chaotic regime known as active turbulence, characterized by the continuous creation and annihilation of topological defects~\cite{sanchez2012,Thampi2013,giomi2013,Ngo2014,Thampi_2014,Shankar2018,Alert2022,Head2024,carenza2025}.

The ability to control the chaotic active nematics dynamics is central to exploiting their potential for applications, as for instance the self-organization, assembly, and transport of suspended passive particles~\cite{Ray2023,Neville2024,Loewe_2022,chandler2024_2,Thampi2016,Nishiguchi2018,Leonardo2009,Reinken2020,carenzaEPL_2020,Houston_2023,VelezCeron2024,kozhukhov2025,Schimming2024}. In the turbulent regime, active nematics create a unique environment where fluid dynamics and defect-mediated transport can be exploited to orchestrate the motion and organization of passive inclusions~\cite{Thampi2016,Carenza2020}. 
Many-body active interactions provide an effective avenue to create new soft composite materials, such as colloidal crystals and self-quenched glasses. Similarly, the behavior of passive soft inclusions, such as lipid vesicles or polymeric droplets, in active nematics could enable novel applications in drug delivery, microfluidics, and programmable self-assembly.
Unlocking this potential requires understanding the interplay between active flows, topology, and passive inclusions. In particular, it is essential to consider how spatial confinement and topological constraints introduced by suspended particles affect emergent behaviors at different activity levels and packing fractions. 

The collective organization of two-dimensional colloidal inclusions in \emph{passive} liquid crystals is strongly influenced by the liquid crystal anchoring~\cite{Stark2001,Smalyuk2005,Smalyuk2018,Genkin2018,Kim2018,senyuk2022}. In the presence of anchoring (either tangential or homeotropic) each suspended particle introduces a local topological charge of $+1$, which the surrounding nematic absorbs by generating a total compensating charge of $-1$. This is typically achieved through the formation of two $-1/2$ topological defects, so that inclusions mutually interact through weak quadrupolar interactions~\cite{Stark2001,Musevich2013}. 
These interactions drive various self-assembled configurations, including linear chains of colloids~\cite{Poulin1997,Lubensky1998}, particle aggregates~\cite{Poulin1998,Yada2004}, defect-stabilized colloidal gels~\cite{Zapoitocky1999}. 
Recent studies suggest that activity can overcome energy barriers and render certain excitations effectively gapless~\cite{Head2024_3}. This opens intriguing possibilities for designing new classes of topological soft materials, where active forces could be harnessed to reshape defect landscapes and guide structure formation beyond equilibrium constraints.

\begin{figure*}[htbp]
\centering
\includegraphics[width=1.0\textwidth]{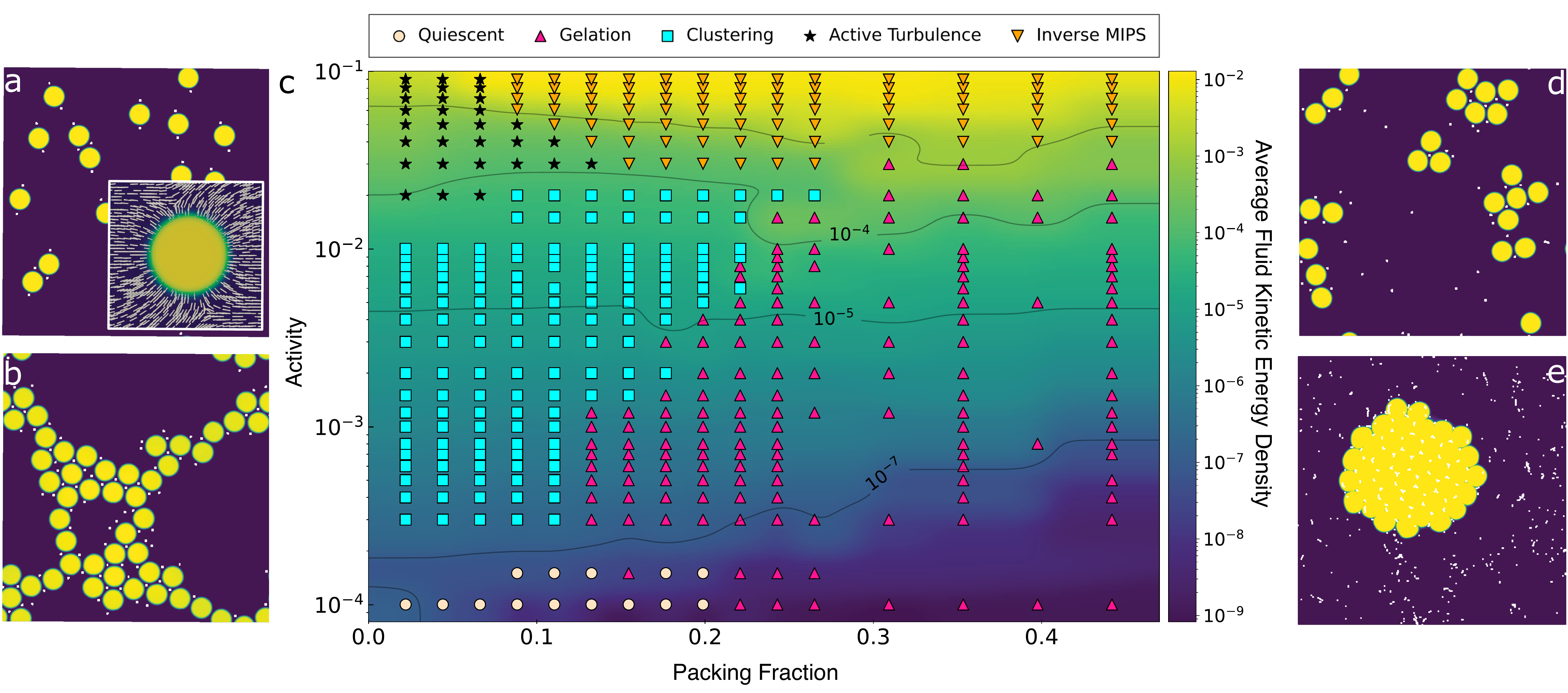}
\caption{\textbf{Morphology and Phase Behaviors of Passive Emulsions in Active Nematics.} Panels~(a, b) and (d, e) illustrate characteristic configurations of the system for different values of activity and packing fraction. Topological defect positions are highlighted in white. (a) Quiescent state, $\zeta=10^{-4}, \Phi=0.09$ ($N=20$);  (b) Gelation state, $\zeta=2\times 10^{-3}, \Phi=0.22$ ($N=50$); (d) Dynamic clustering state, $\zeta= 4 \times 10^{-3}, \Phi=0.11$ ($N=25$); (e) Fully clustered state, $\zeta= 7 \times 10^{-2}, \Phi=0.18$ ($N=40$). The inset in panel~a depicts the director field in proximity of a droplet. (c) Phase diagram showing the system's morphological behavior as a function of packing fraction $\Phi$ and activity  $\zeta$. The background color represents the average kinetic energy density of the fluid, as indicated by the color bar.}  
\label{fig1}
\end{figure*} 

While active droplets suspended in passive fluids~\cite{carenza2019,Mullol2020,Guillamat2018,Singh2020,carenzaphysicaAchol,tjhung2017,Copar2019,Nejad2023,Negro2025,Alam2024} and colloidal suspensions in passive liquid crystals~\cite{Lavrentovich2014,Tkalec2013,head2024_2,Copar2015,Luo2018,Wamsler2024,Chandler2024} have been extensively studied, the case of passive inclusions suspended in active nematics has received far less attention.
Recent research demonstrated that a single passive inclusion suspended in an active nematic exhibits anomalous diffusivity, deviating significantly from Brownian motion due to the interaction between bulk defects and the inclusion-induced nematic deformations~\cite{singh2024,kozhukhov2025}. This behavior is strongly dependent on the ratio of the droplet radius to the active length scale, with smaller droplets experiencing significant drifts driven by active flows. 

However, whether this behavior can be harnessed for material self-assembly remains an open question, requiring systematic investigation of the dynamical and morphological properties in dense droplet suspensions.
In this study, we explore the full phase diagram of a system of multiple passive soft droplets, modeled as multiphase fields, suspended in an active nematic fluid. We systematically vary both the activity level and the droplet packing fraction, revealing a rich diversity of dynamical and morphological regimes.
At intermediate activity levels, we find a dynamic clustering behavior also observed in hard colloidal suspensions~\cite{foffano2019}. At higher packing fractions, the clustered phase gives way to a gel-like phase where droplets organize into a percolating network spanning the entire system. Crucially, when activity exceeds the threshold for active turbulence, the system undergoes a striking transition: the suspended droplets completely cluster, exhibiting tunable diffusive and morphological properties. This regime resembles motility-induced phase separation (MIPS) observed in systems of self-propelled particles~\cite{marchetti,fily2012,gonnellaMIPS,Semeraro_2024,rednerMIPS,buttSPP,cates2015}, but with a fundamental distinction. In classical MIPS, phase separation arises from the particles' own activity, whereas in our system, it is driven by the motility of the active solvent while the suspended droplets remain passive. 
Accordingly, we term this regime \emph{inverse} MIPS.

\begin{figure*}[htbp]
\centering
\includegraphics[width=1.0\textwidth]{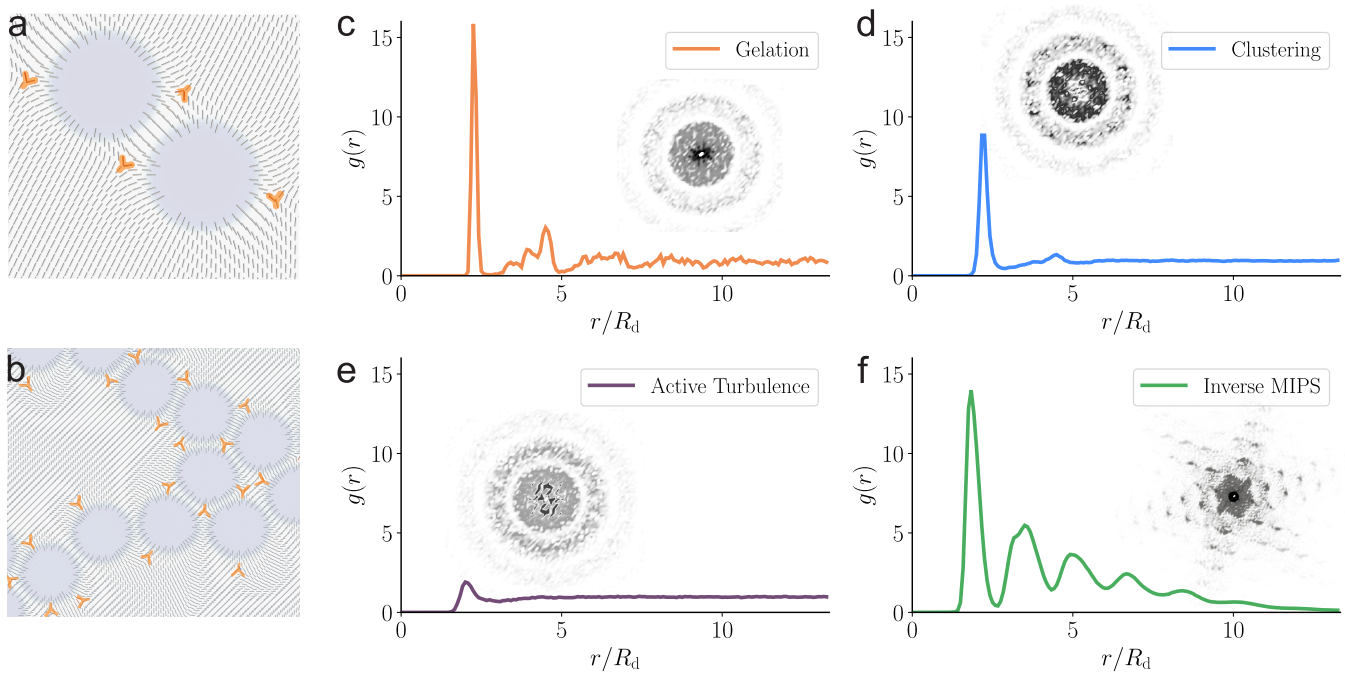}
\caption{\textbf{Morphological Characterization of the Passive Emulsion.}  
Panels (a, b) show representative configurations of the director field in the clustered and gelation phase, respectively. The radial distribution function \(g(r)\) for each phase is shown in panels (c--f). Insets in these panels display the corresponding two-dimensional structure factor \(\mathcal{S}_\phi\).} 
\label{fig2}
\end{figure*}

\subsection*{Results}

\paragraph*{Morphological Characterization.}

To investigate the collective behavior of passive soft 
droplets suspended in an active nematic fluid, we employ a multiphase field approach. Each droplet is represented by a distinct phase-separating field $\phi_i$, where $i$ uniquely identifies the droplet. The orientational state of the active nematic fluid is described by the $Q$-tensor, defined as $\bm{Q} = S(\bm{n}\bm{n} - \bm{I}/2)$. Here, $S$ represents the nematic order parameter, quantifying local nematic order, and $\bm{n}$ is the head-tail symmetric unit vector indicating the local alignment direction of nematic molecules.

We numerically solve the continuum active nemato-hydrodynamic equations in a two-dimensional domain with periodic boundary conditions, coupled to the conserved dynamics of the suspended passive droplets. This model has previously been validated for reproducing the dynamics of active suspensions confined on rigid and deformable shells, active liquid interfaces, and for predicting patterns in three-dimensional active suspensions. [See the Methods Section for a detailed description of the numerical model.]

Non-equilibrium energy injection is introduced in the Navier-Stokes equation via the active stress tensor $\bm{\sigma}^{\rm act} = -\zeta \bm{Q}$, where $\zeta$ is the activity parameter. At low activity levels ($\zeta \lesssim 2 \times 10^{-4}$), the system remains in a quiescent state dominated by passive effects. When the liquid crystal exhibits homeotropic or tangential anchoring at the droplet interface, each droplet becomes topologically equivalent to a $+1$ defect. To compensate, the liquid crystal generates a pair of semi-integer negative defects at antipodal points (Fig.~\ref{fig1}a). This configuration induces a quadrupolar short-range interaction, too weak to drive droplet clustering. The resulting structure depends on the initial conditions and droplet packing fraction $\Phi$. Specifically, droplets either remain near their initial positions (\href{https://youtube.com/shorts/tsXMZsQl3Ts?feature=share}{Movie~S1}) or form a percolating network (Fig.~\ref{fig1}b)  when $\Phi$ exceeds a critical value of $\Phi^{*} = 0.21$, as shown in the phase diagram in Fig.~\ref{fig1}c.

\begin{figure*}[htbp]
\centering
\includegraphics[width=1.0\textwidth]{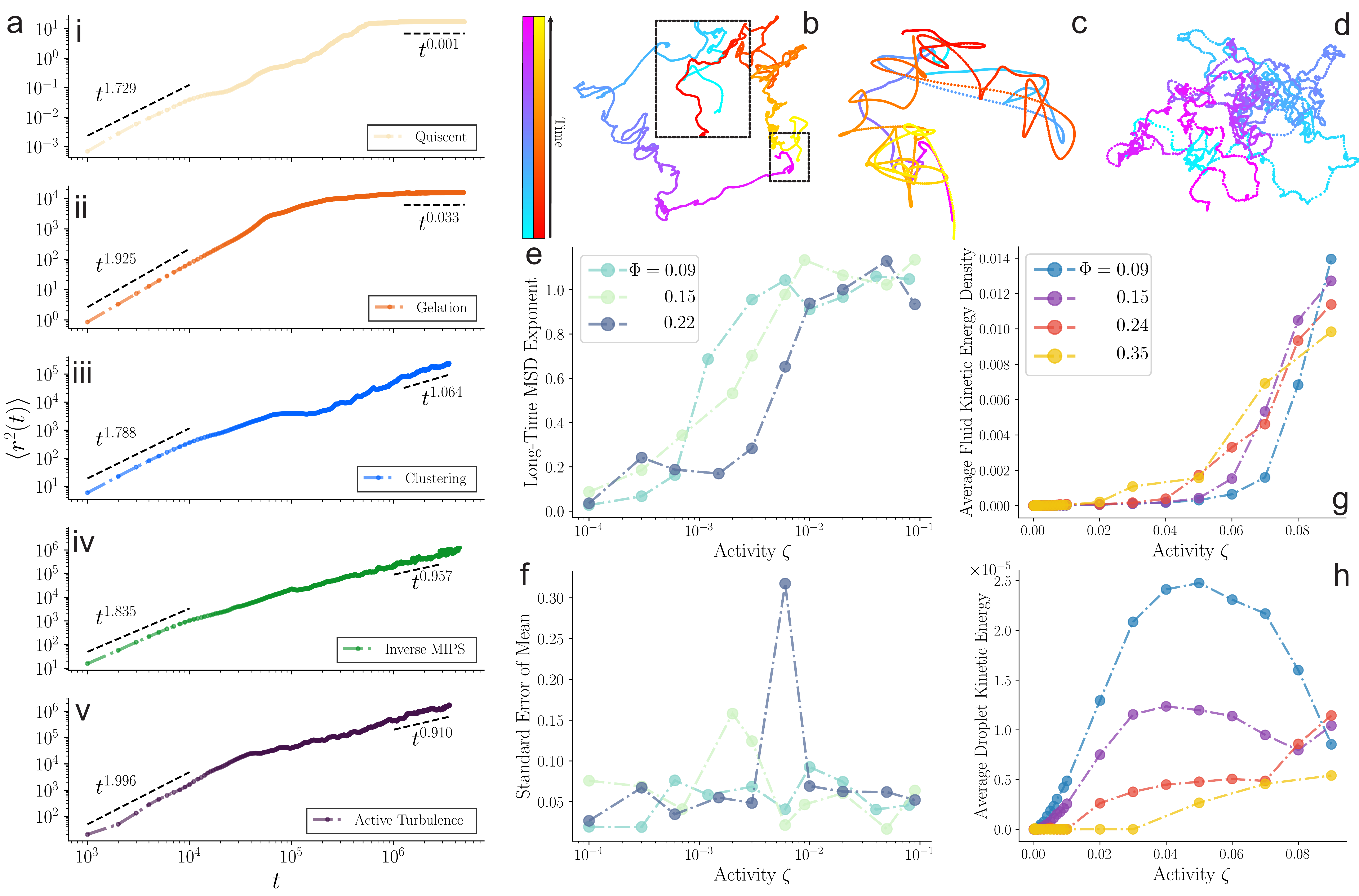}
\caption{\textbf{Kinetic Characterization of the Passive Emulsion.} (a) Mean square displacement (MSD) averaged over multiple droplets in the system for various phases. Dashed lines indicate best-fit slopes for short- and long-term dynamics. (b–d) Representative droplet trajectories in (b) the clustering phase, (c) the inversed MIPS phase, and (d) the active turbulence regime.  (e) Long-time MSD exponent as a function of activity for three packing fractions, obtained by averaging over $n = 5$ independent simulations. (f) Standard error of the mean for the long-time MSD exponent at varying activity levels, for the same packing fraction as in panel~e. (g) Average droplet kinetic energy versus activity, for different packing fractions. (h) Average fluid kinetic energy density versus activity, for the same packing fractions as in panel~g.} 
\label{fig3}
\end{figure*}

\begin{figure*}[htbp]
\centering
\includegraphics[width=0.9\textwidth]{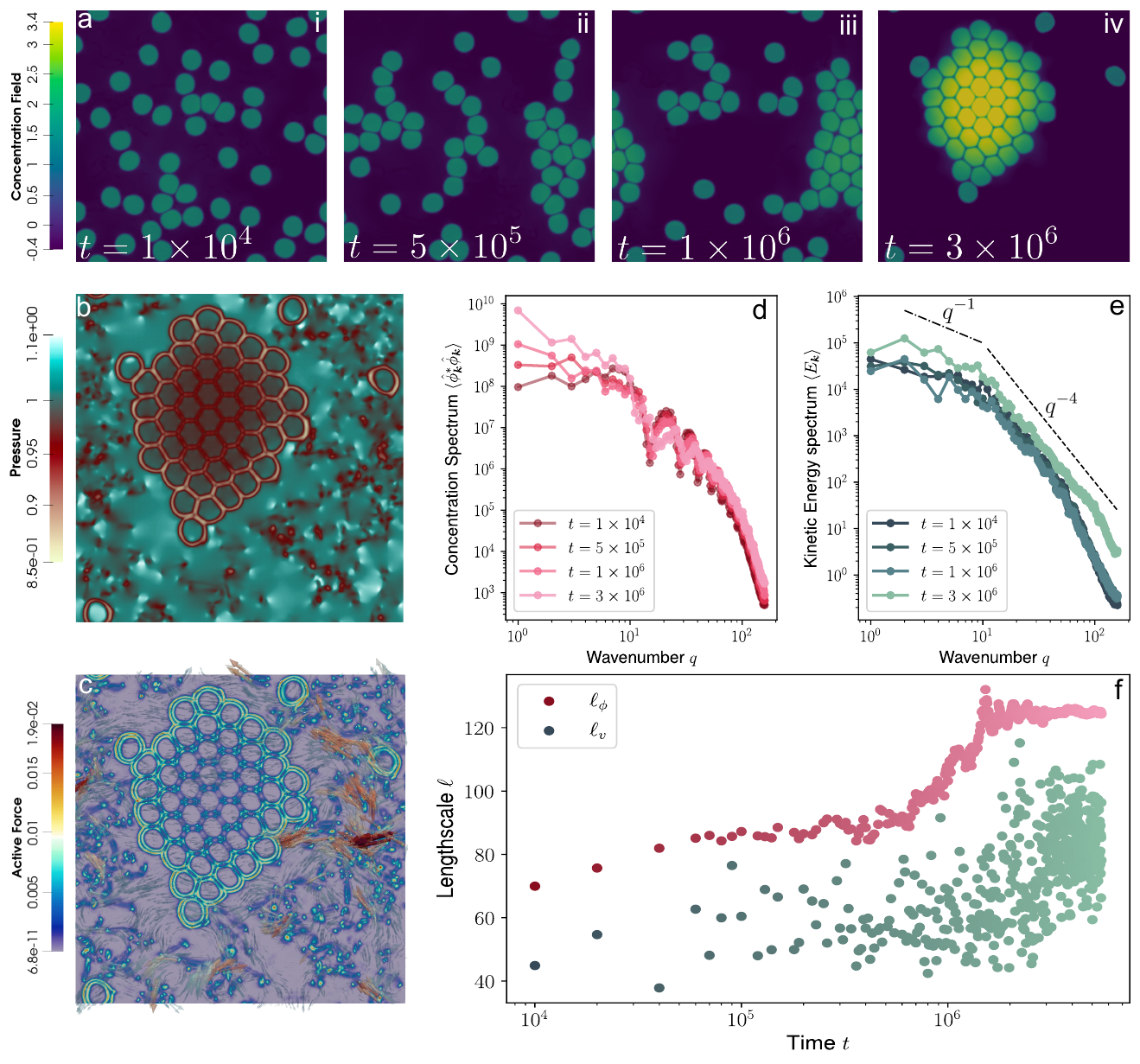}
\caption{\textbf{Clustering Dynamics.}  
(a) Time-lapse sequence of the system configuration for $N = 50$ and $\zeta = 0.06$, resulting in a fully separated state. (b--c) Color maps of the hydrodynamic pressure (panel b) and the active force (panel c) at time $t = 3 \times 10^{-6}$. The velocity field, displaying multiple vortices outside the cluster, is superimposed on the color map in panel c. (d--e) Temporal evolution of the structure factor of the concentration field (panel d) and the kinetic energy spectrum (panel e). (f) Characteristic cluster size $\ell_\phi$ and flow structure size $\ell_v$---derived as the inverse of the first moment of their respective structure factors---plotted versus time. The color scale corresponds to that used in panels (d, e).} 
\label{fig4}
\end{figure*}

A dramatic transition occurs when activity exceeds the threshold for spontaneous flow ($\zeta \gtrsim 2 \times 10^{-4}$). Droplets are advected by active jets and undergo a \emph{stop and go} motion~\cite{singh2024}.
However, in contrast to the case of isolated droplets, the dynamics and collective arrangement in a crowded environment are significantly altered by their mutual interaction, mediated by the active liquid crystal.
Specifically, when two droplets approach each other, anchoring-induced radial deformations in the surrounding liquid crystal stabilize the formation of \emph{defect bonds}. These bonds arise when droplets share two (or more) of their surrounding topological defects, causing them to adhere (see Fig.~\ref{fig2}a,b). Such droplet dimers act as aggregation centers, facilitating the growth of larger clusters through the absorption of isolated droplets or the merging with other clusters.

Interestingly enough, at intermediate activity levels $(\zeta \lesssim 10^{-2}$), cluster growth is arrested due to the disruptive effects of active jets, which undermine the structural stability of the clusters (\href{https://youtube.com/shorts/mcLeJ4XFR68?feature=share}{Movie~S2}). This behavior is prevalently observed at low packing fractions and intermediate values of activity (Fig.~\ref{fig1}c,d).
Consequently, the system exhibits highly dynamic behavior, with clusters continuously assembling and disassembling over time.
To morphologically characterize the system, we show in Fig.~\ref{fig2}c-f the averaged droplet radial distribution function $g(r)$ (see Methods) for different morphological regimes.
In the clustering phase, Fig.~\ref{fig2}d reveals a pronounced peak corresponding to the first-neighbor contribution, along with a smaller second-order peak, consistent with the finite size of the cluster. The two-dimensional structure factor $\mathcal{S}_{\phi}=\langle \phi^{*}(\bm{k}) \phi(\bm{k}) \rangle$ (with $\phi = \sum_i \phi_i$ the global concentration field) further exhibits broken rotational symmetry, reflecting the local arrangement of droplets within the clusters.

\begin{figure*}[htbp]
\centering
\includegraphics[width=0.8\textwidth]{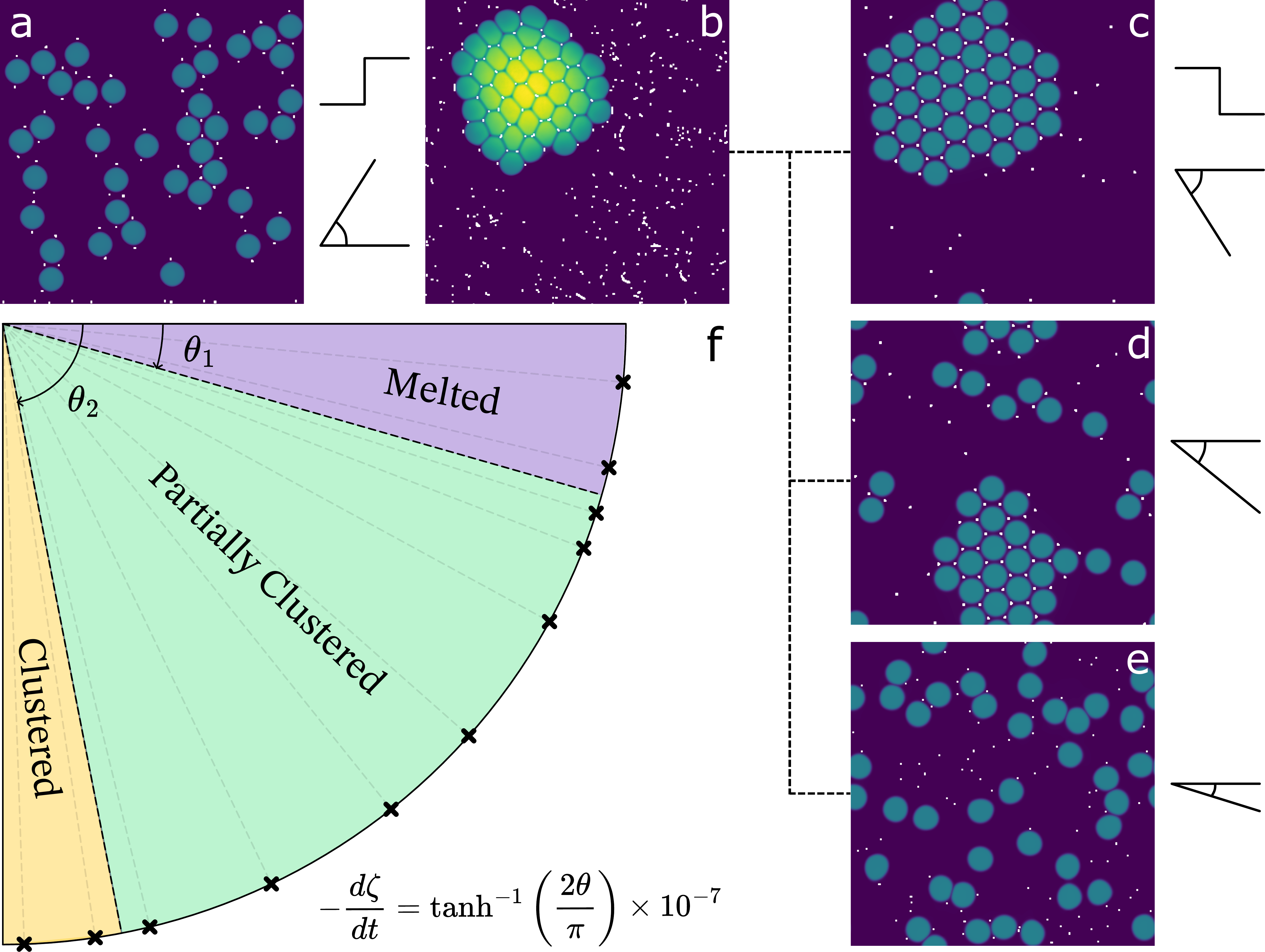}
\caption{\textbf{Rate-Dependent Hysteresis of Full Clustering.} Panels (a,b) and (d-f) illustrate different configurations of the system for $\Phi=0.18$ $(N=40)$. Topological defect positions are highlighted in white. (a) Randomly initialized passive configuration; (b) Fully clustered state, by increasing activity until $\zeta=7 \times 10^{-2}$, either abruptly or linearly. Panels (d,e,f) illustrate characteristic final passive configurations of the system for different values of linear decrease rate of activity. (d) Clustered state (abrupt reduction);  (e) Partially clustered state (intermediate reduction rate);  (f) Melted state (slow reduction rate). (c) Angular diagram showing the system's morphological behavior as a function of $\theta \equiv \frac{\pi}{2} \tanh{\left(-\frac{d\zeta}{dt} \times 10^7 \right)}$. Two transitions occur at $\theta_1 \approx 16^\circ$ $\left(-\frac{d\zeta}{dt}\approx1.8\times10^{-8}\right)$ and $\theta_2 \approx 79^\circ$ $\left(-\frac{d\zeta}{dt}\approx1.4\times10^{-7}\right)$.}
\label{fig5}
\end{figure*} 

When the activity is increased beyond  $\zeta \gtrsim 10^{-2}$ at low packing fractions ($\Phi \lesssim 0.15$), the active flow becomes large enough to prevent clustering (\href{https://youtube.com/shorts/mi0Bww6evIU?feature=share}{Movie~S4}), leading the system to transition into a new regime characterized by chaotic droplet motion and the absence of structures (isotropic concentration spectrum and flat $g(r)$ in Fig.~\ref{fig2}e). We identify this state as \emph{active turbulence}, where the continuous active injection destabilizes the liquid crystal, resulting in deformed patterns with a characteristic length scale of  $\ell_{\rm act} = \sqrt{L/\vert\zeta\vert}$, where $L$ represents the elastic stiffness of the liquid crystal. 
In this regime, the liquid crystal relaxes the elastic stress caused by activity-induced deformations by nucleating pairs of topological defects. This process leads to the proliferation of topological defects throughout the system, further contributing to the active turbulence dynamics.

Remarkably, at very high activity levels, we observe droplets reorganizing into clusters once again, despite the presence of strong active jets (\href{https://youtube.com/shorts/e-iF8TSjn9U?feature=share}{Movie~S5}). While defect bonds are primarily responsible for cluster stability at lower activity levels (light blue points in the phase diagram of Fig.~\ref{fig1}a), at high activity, the pressure exerted by the active liquid crystal becomes sufficiently large to stabilize the clusters--a mechanism which is reminiscent of an \emph{inverse} motility-induced phase separation (orange points in Fig.~\ref{fig1}a). 
As a result, these clusters remain fully separated, grow over time, and retain their constituent droplets without loss. This occurs because the steric constraints imposed by the surrounding active nematic matrix effectively trap the droplets in place, arranging them in hexatic fashion as shown by the $g(r)$ and $2D$ structure factor in Fig.~\ref{fig2}f.

The existence of this fully clustered state at high activity is one of the most significant findings reported in this article. Notably, this behavior is largely independent of the packing fraction of the system, highlighting the robustness of steric trapping at large activity. 
Finally, as illustrated in the  phase diagram in Fig.~\ref{fig1}c, a transition from a percolating network to a fully separated phase also occurs at large packing fractions.

\paragraph*{Kinetic Characterization.}

Fig.~\ref{fig3}a presents the mean square displacement (MSD) $\langle r^2(t) \rangle$ of droplets across different phases. In the quiescent and gelation phases at low activity, droplets reach a stationary configuration at long times (Fig.~\ref{fig3}a(i-ii)), while in the three higher-activity regimes, they exhibit long-time diffusive behavior (Fig.~\ref{fig3}a(iii-v)). Notably, this diffusive dynamics emerges independently of packing fraction, as demonstrated by the behavior of the long-time MSD exponents versus activity (Fig.~\ref{fig3}e), obtained by performing statistical average over independent simulations for different packing fractions (see Methods section).  

In general, the system undergoes a transition from subdiffusive behavior at low activity to diffusive behavior at larger activity, with the packing fraction influencing the specific activity required for this transition; higher packing fractions reduce the available space for the active nematic instability, necessitating greater activity to generate self-sustained flows and sustain droplet diffusive motion. 
Interestingly enough, near the transition between gelated and clustered states, we observe significant fluctuations in the MSD exponents (Fig.~\ref{fig3}f), evidenced by peaks in the standard error of the mean. These fluctuations suggest phase coexistence during the transition to the fully clustered phase at high activity, whereas the transition to the turbulent state at lower packing fractions appears continuous.

Interestingly, the kinetic energy of the droplets $E_d = \langle m_d \dot{\bm{r}}^2/2 \rangle $ and the fluid $E=\langle \rho v^2/2 \rangle $ exhibit distinct behaviors as the activity level is varied. 
While the fluid's kinetic energy increases monotonically with activity (Fig.~\ref{fig3}g), the droplets' kinetic energy shows a more complex behavior (Fig.~\ref{fig3}h). At low packing fractions $\Phi<0.27$, $E_d$ first increases with activity during the onset of spontaneous motion, then plateaus in the turbulent regime, and finally drops in the inverse MIPS regime at large activity. This non-monotonic behavior can be rationalized in terms of the changing flow structure at increasing activity. At small activity values droplet advection is enhanced by self-sustained flows. Conversely, at larger activity values the typical size of the uncorrelated active vortices is significantly smaller than the cluster size (see Fig.~\ref{fig4}c for instance). This reduces the advective effect of the turbulent active fluid, eventually causing a drop in $E_d$ despite the increasing activity .

\paragraph*{Clustering Dynamics.}

The dynamics of inverse MIPS reveals a complex interplay between active flows and passive droplet organization. As illustrated in Fig.~\ref{fig4}a, the process begins with the formation of small, transient clusters (panels a(i-ii)) similar to those observed at lower activity levels. However, at these elevated activity values, the system exhibits a qualitatively different behavior as clusters progressively merge into larger and more stable structures (panels a(iii-iv)). These emergent superstructures serve as nucleation centers that incorporate smaller clusters and individual droplets through an ongoing coarsening process. Notably, these macroscopic clusters eventually reach a stationary state where the internal droplets are organized into hexatic configurations.

The stabilization mechanism of these clusters differs fundamentally from the defect-mediated bonding observed at lower activities. In this high-activity regime, where the interstitial space between droplets becomes depleted of liquid crystal, hydrodynamic effects dominate the collective behavior. Cluster stabilization arises from the pressure gradient that develops as a consequence of asymmetric forces acting on peripheral droplets. While interior droplets experience balanced interactions with their neighbors, peripheral droplets are subjected to strong active stresses from the surrounding turbulent fluid. This imbalance establishes a radially decreasing hydrodynamic pressure profile $P_{\rm h}$ toward the cluster center, as clearly evidenced in Fig.~\ref{fig4}b-c.

The dynamics of inverse MIPS we observe here bears significant similarity to conventional MIPS observed in active colloidal systems, following an analogous two-stage progression. In both cases, phase separation results from a density-dependent mobility reduction. In more detail, in our case, passive droplets transported by active flows experience decreased mobility upon collisions and elastic interactions which eventually lead to local accumulation. This process is complemented by a feedback mechanism in which accumulating droplets progressively exclude the active fluid from the cluster interior. The expelled active fluid then exerts compressive stresses on the cluster periphery, which are transmitted inward through the compression of the droplet network. This explains the crucial role of droplet compressibility in cluster stability [our observations indicate that increasing surface tension--and thus reducing compressibility--significantly destabilizes droplet clusters].

The temporal evolution of the system's organization is quantitatively captured through spectral analysis. The concentration structure factor $\langle \hat{\phi}^*_{\bm{k}} \hat{\phi}_{\bm{k}} \rangle$ (Fig.~\ref{fig4}d) reveals two important features: the development of Bragg peaks at large wavenumbers reflects the emergence of hexatic order within clusters, while the growing intensity at small wavenumbers indicates the progressive increase in cluster size. Furthermore, the kinetic energy spectrum $E_k = \langle \hat{\bm{v}}^*_{\bm{k}} \cdot \hat{\bm{v}}_{\bm{k}} \rangle$ (Fig.~\ref{fig4}e) maintains the characteristic scaling of active turbulence, demonstrating that the active fluid retains its chaotic nature despite the presence of the separated cluster. From these spectra, we extract the characteristic length scales of the concentration field $\ell_\phi$ and velocity field $\ell_v$ as the first moments of their respective distributions. As shown in Fig.~\ref{fig4}f, $\ell_v<\ell_\phi$ throughout the evolution, confirming that the turbulent vortices remain smaller than the emergent droplet structures. Such a length-scale separation highlights the hierarchical organization of the system, where microscopic active turbulence coexists with macroscopic droplet ordering.

\subsection*{Discussion}
Understanding how clustered and inverse MIPS states respond to time-dependent activity changes is crucial for developing robust self-assembly strategies. To investigate this dynamic response, we conducted systematic numerical experiments that probe the system's behavior under  different activity reduction rates.
Fig.~\ref{fig5} shows the results of this investigation. A disordered configuration with packing fraction $\Phi=0.18$ at the initial time (panel~a) is evolved by increasing activity until $\zeta = 7 \times 10^{-2}$, to obtain a fully clustered state through inverse MIPS (panel~b).
Starting from this configuration we then proceed to reduce the activity at different rates. We find three possible regimes.
When activity is rapidly switched-off the system maintains its clustered configuration through persistent defect-mediated bonds, with the interstitial nematic forming an hexatic matrix of defects between droplets (panel~c and \href{https://youtube.com/shorts/SGwrGd3eXkY?feature=share}{Movie~S6}). 

More gradual activity reduction leads to different morphological behaviors. At intermediate decay rates, we observe the fully separated cluster melt, as decreasing active pressure renders peripheral droplets vulnerable to the advective action of turbulent flows. This creates an intriguing coexistence regime where the initial cluster persists alongside a dispersed droplet suspension (panel~d and \href{https://youtube.com/shorts/NsRfsqugN5E?feature=share}{Movie~S7}). The slowest activity reduction rates ultimately restore complete disorder, with the system adiabatically returning to its initial configuration through complete melting (panel~e and \href{https://youtube.com/shorts/ruMld3cctlc?feature=share}{Movie~S8}).

These rate-dependent pathways, summarized in the phase diagram of Fig.~\ref{fig5}f, establish temporal activity modulation as a powerful tool for designing reconfigurable activity-driven composites. The ability to select between arrested, partially melted, or fully disordered states through control of activity dynamics suggests new possibilities for designing materials with tunable structural memory. 

From a broader perspective, our work demonstrates how active nematics can orchestrate the behavior of embedded soft particles through the interplay of hydrodynamic flows and topological constraints. The observed transitions—from turbulence-driven clustering to gelation and inverse MIPS—significantly broaden the toolkit for non-equilibrium self-assembly by harnessing the chaotic dynamics of active turbulence. We anticipate these findings will stimulate experimental efforts using both biological and synthetic active nematics, particularly to explore how droplet softness, anchoring conditions, and three-dimensional confinement influence the reported phenomenology. 
The principles revealed in this study could ultimately guide the design of adaptive smart materials capable of programmable structural transitions.

\FloatBarrier

\section*{Material and Methods}

\subsection*{The model}
We consider a suspension of isotropic droplets in a two-dimensional active nematic liquid crystal. Each droplet is described using a multiphase approach, modeled by phase fields $\phi_i({\bm r}, t)$, with $i=1,\dots,N$, where $N$ represents the total number of droplets in the system. The nemato-hydrodynamics is characterized by the incompressible velocity field $\bm{v}({\bm r},t)$, constrained by $\nabla \cdot \bm{v} = 0 $, and the nematic $Q-$tensor ${\bm Q}({\bm r},t) = S(\bm{n}\bm{n}-\bm{I}/2)$. Here, the principle eigenvector $\bm{n}$ --the so-called \emph{director field}-- defines the local direction of alignment of the liquid crystal, while its magnitude $S$ quantifies the degree of local nematic order.

\paragraph*{Evolution equations.} The dynamics of the fields is governed by the following set of coupled PDEs,
\begin{eqnarray}
&D_t \phi_i = M \nabla^2 \mu_i \; , \label{eqn:2} \\
&D_t{\bm Q}= {\bm S}({\bm W},{\bm Q}) + \gamma^{-1} {\bm H} \; , \label{eqn:3} \\
&\rho D_t {\bf v} = \nabla\cdot ({\bm \sigma}^{\rm hydro}+{\bm \sigma}^{\phi}+{\bm \sigma}^{\rm LC}+{\bm \sigma}^{\rm act}) \; , \label{eqn:4} 
\end{eqnarray}
where the operator $D_t= \partial_t + \bm{v}\cdot \nabla$ denotes the material derivative. Eq.~\eqref{eqn:2} is the Cahn-Hilliard equation for the conserved concentration fields $\phi_i$, with $M$ the mobility parameter, and $\mu_i = \delta \mathcal{F}/\delta \phi_i$ the chemical potentials of each droplet, $\mathcal{F}$ being the free energy of the system introduced below.
Eq.~\eqref{eqn:3} is the Beris-Edwards equation ruling the dynamics of the $Q-$tensor. The operator 
\begin{multline}
\bm{S}({\bm W},{\bm Q})=(\xi{\bm D}+{\bm \Omega})({\bm Q}+{\bm I}/3)\\
+ (\xi{\bm D}-{\bm\Omega})({\bm Q}+{\bm I}/3) \\
-2\xi({\bm Q}+{\bm I}/3) {\rm Tr} ({\bm Q}{\bm W}) \; ,
\label{eqn:5}
\end{multline}
denotes the co-rotational derivative and defines the dynamical response of the LC to straining and shearing.
Here, ${\bm D}=({\bm W}+{\bm W}^T)/2$ and ${\bm\Omega}=({\bm W}-{\bm W}^T)/2$ are the symmetric and anti-symmetric part of the velocity gradient tensor $W_{\alpha\beta}=\partial_{\beta}v_{\alpha}$, respectively. The flow alignment parameter $\xi$ determines the aspect-ratio of the LC molecules and the dynamical response of the LC to an imposed shear flow. Here, we choose $\xi = 0.7$ to consider flow-aligning rod-like molecules. 
The coefficient $\gamma$ in Eq.~\eqref{eqn:3} is the rotational viscosity measuring the relative importance of advection with respect to relaxation, and the molecular field ${\bm H}=-\frac{\delta {\mathcal F}}{\delta {\bm Q}}+({\bm I}/3){\rm Tr}\frac{\delta {\mathcal F}}{\delta {\bm Q}}$.
Finally, Eq.~\eqref{eqn:4} is the Navier-Stokes equation for the incompressible velocity field ($\nabla \cdot \bm{v} = 0$) with constant density $\rho$. Here, the stress tensor has been divided in: \emph{(i)} a hydrodynamic contribution ${\bm \sigma}^{\rm hydro}= - P \bm{I} + \eta \nabla \bm{v}$ where $P$ is the hydrodynamic pressure ensuring incompressibility, and $\eta$ is the shear viscosity; \emph{(ii)} an interface contribution
\begin{equation}
\sigma^{\phi}_{\alpha\beta}=\sum_{i=1}^N\left[\left(f-\phi_i\frac{\delta{\cal F}}{\delta\phi_i}\right)\delta_{\alpha\beta}-\frac{\partial f}{\partial(\partial_{\beta}\phi_i)}\partial_{\alpha}\phi_i\right] \; ;
\label{eqn:6}
\end{equation}
where $\mathcal{F} \equiv \int_V d\bm{r} \ f$ is the free energy; \emph{(iii)} a LC contribution 
\begin{eqnarray}
\sigma_{\alpha\beta}^{LC}=&&-\xi H_{\alpha\gamma}(Q_{\gamma\beta}+\frac{1}{3}\delta_{\gamma\beta})-\xi(Q_{\alpha\gamma}+\frac{1}{3}\delta_{\alpha\gamma})H_{\gamma\beta}\nonumber\\
&&+2\xi(Q_{\alpha\beta}-\frac{1}{3}\delta_{\alpha\beta})Q_{\gamma\mu}H_{\gamma\mu}+Q_{\alpha\gamma}H_{\gamma\beta}-H_{\alpha\gamma}Q_{\gamma\beta}\nonumber\\
&&-\partial_{\alpha}Q_{\gamma\mu}\frac{\partial f}{\partial(\partial_{\beta}Q_{\gamma\mu})} \; ,
\label{eqn:7}
\end{eqnarray}
accounting for the elastic backflow; and \emph{(iv)} the active stress $\bm{\sigma}^{\rm act} = -\zeta \bm{Q}$, where $\zeta$ is the activity parameter, positive for extensile systems.

\paragraph*{Free Energy.}
The equilibrium properties of the system are defined by the free energy
\begin{equation}
    \mathcal{F} = \int_V d\bm{r} \ f = \int_V \ d\bm{r} \ \left( f^{\phi} +  f^{\rm LC} + f^{\rm anch} \right) \; , 
    \label{eqn:1}
\end{equation}
where $f^{\phi}$ is the emulsion free energy density, $f^{\rm LC}$ the liquid crystal free energy accounting for the isotropic-nematic transition and elastic properties of the LC, and $f^{\rm anch}$ defines the anchoring properties of the LC at the droplets' surface. 
In particular,
\begin{equation}
   f^{\phi} = \sum_{i=1}^{N} \frac{a}{4}\phi_i^2(\phi_i-\phi_0)^2 + \frac{k_\phi}{2}\sum_{i=1}^{N}(\nabla\phi_i)^2  + \sum_{i,j,i<j}\epsilon\phi_i^2\phi_j^2 \; .
   \label{eqn:A1}
\end{equation}
For $a,k_{\phi}>0$, $f^{\phi}$ describes phase separating fields with two minima at $\phi_i=\phi_0$ and $\phi_i=0$ that, at equilibrium, arrange into circular $2D$ droplets with surface tension $\sigma=\sqrt{8ak_{\phi}/9}$, and interface thickness $\xi_{\phi} = \sqrt{2k_{\phi}/a}$. The interaction term proportional to $\epsilon>0$ describes soft repulsion pushing droplets apart when overlapping.

The LC free energy density
\begin{equation}
    \begin{split}
    f^{\rm LC } = A_0\left[ \dfrac{1}{2} \left(1-\frac{\chi(\phi)}{3}\right) {\rm Tr} {\bm Q}^2 -\frac{\chi(\phi)}{3} {\rm Tr} {\bm Q}^3 \right. \\ \left. +\frac{\chi(\phi)}{4} \left( {\rm Tr} {\bm Q}^2 \right)^2 \right]+\frac{L}{2} (\nabla \cdot \bm{Q})^2 \; .
    \label{eqn:A2}
 \end{split}
\end{equation}
Here, the bulk constant $A_0>0$, and $\chi(\phi)$ is a temperature-like parameter that drives the isotropic-nematic transition that occurs for $\chi(\phi)>\chi_{cr}=2.7$~\cite{degennes}, where we have denoted with $\phi=\sum_i \phi_i$ the total concentration field. We choose $\chi=\chi_0+\chi_s\phi$, with  $\chi_0>2.7$ and $\chi_s<0$ to confine the LC outside of the droplet where $\phi=0$ (see Fig.~\ref{fig1}a). The last term in Eq.~\eqref{eqn:A2} captures the energy cost of elastic deformations in the single elastic constant approximation. 

Finally, the anchoring contribution to the free-energy
\begin{equation}
     f^{\rm anch } = W \sum_{i=1}^N \left[\partial_{\alpha}\phi_i Q_{\alpha\beta} \partial_{\beta}\phi_i\right]. 
     \label{eqn:A3}
\end{equation}
defines the orientation of the LC at the shell's interface. For $W<0$, ($W>0$) homeotropic (tangential) anchoring is imposed at the droplets' surface.

\subsection*{Simulation protocol}
We integrate the dynamics of the hydrodynamic fields in Eq.~\eqref{eqn:2}-\eqref{eqn:4} in a square grid of size $L=320$ with a predictor-corrector hybrid lattice Boltzmann approach~\cite{succi2018,Negro2023,carenza2019,Carenza2019_2} consisting of solving Eq.~(\ref{eqn:2},\ref{eqn:3})  with a finite-difference algorithm implementing first-order upwind scheme and fourth-order accurate stencils for space derivatives, and the Navier-Stokes equation (Eq.~\eqref{eqn:4}) through a predictor-corrector LB scheme on a $D2Q9$ lattice. For technical details on the lattice Boltzmann method here implemented refer to References~\cite{Negro2023}.

The system is initialized as follows. The concentration field of every droplet $\phi_i$ is set up as a spherical droplet with radius $R$, where $\phi_i=\phi_0$ inside the droplet and $\phi_i=0$ outside it. 
The $Q-$tensor is initialized such that $S=0$ inside the droplets, while outside is initialized with a small amplitude and a random orientation. The velocity field is initialized to $0$.

The key control parameter of the system are: \emph{(i)} the droplets' packing fraction $\varPhi = N \pi R^2/L^2$; and \emph{(ii)} dimensionless activity number 
\begin{equation}
    A = \dfrac{\ell^2_{\rm c}}{\ell^2_{\rm act}} =  \dfrac{\vert \zeta \vert }{A_0}\; ,
    \label{eqn:10}
\end{equation}
where $\ell_{\rm act}=\sqrt{L/\vert \zeta \vert}$ is the active length-scale.

\paragraph*{Simulation parameters}
We have varied two control parameters: \emph{(i)} the packing fraction $\varPhi$ was varied between 0.02 and 0.44, by considering systems comprising a different number of droplets, namely $N=5$, up to $N=100$; \emph{(ii)} the activity $\zeta$ was varied between $0$ and $9\times 10^{-2}$. 
The other model parameters are chosen as follows. We fixed the droplet radius to $R=12$ and mobility parameter to $M=0.1$. The free energy parameters are $a=0.07,\,\phi_0=2.0,\,k_\phi=0.1,\,\epsilon=0.05,\,A_{0}=0.2,\,L=0.01,$ and $W=-0.04$. To confine LC outside of the droplets, the parameters $\chi_0=3.0$ and $\chi_s=-0.25$ are selected. All simulations were performed at a fixed system size of $L=320$, where we used periodic boundary conditions along both directions.

\paragraph*{Statistical analysis of long-time MSD exponent}
To investigate the long-time dynamical behavior, we analyze the MSD across a range of packing fractions and activity levels. Specifically, we consider three packing fractions: $\Phi=0.09, 0.15,0.22$. For each packing fraction, ten distinct activity levels are considered. At each combination of packing fraction and activity level, we perform $n=5$ independent simulations.

For every simulation, the MSD is computed as a function of time. A power-law fit of the form $\text{MSD}(t) \sim t^\alpha$ is applied in the long-time regime to extract the MSD exponent $\alpha$. The exponents obtained from the five independent simulations are then used to compute the mean and the standard error of the mean (SEM) plotted in Fig~\ref{fig3}e and Fig.~\ref{fig3}f, respectively.

\subsection*{Measurements}

\paragraph*{Mean squared displacement analysis.} The mean squared displacement (MSD) at a time $t$ is defined as follows
\begin{equation}
    \left\langle \left| \mathbf{r}(t) - \mathbf{r}_0 \right|^2 \right\rangle = 
    \frac{1}{N} \sum_{i=1}^{N} \left| \mathbf{r}^{(i)}(t) - \mathbf{r}^{(i)}(t_0) \right|^2\; ,
\label{eqnA7}
\end{equation}
where $N$ is the number of droplets, $\mathbf{r}^{(i)}(t)$ is the position vector of the $i$-th droplet's center of mass at time $t$, and $t_0$ is the reference time for the initial positions of droplets.

\paragraph*{Radial distribution function.} 
The radial distribution function $g(r)$ describes the probability of finding a particle at a distance $r$ from a reference particle. It is defined as:
$
g(r) = \frac{1}{\rho_0 N} \left\langle \sum_{i \neq j} \delta(r - r_{ij}) \right\rangle,
$
where $\rho_0=N/L^2$, \,$r_{ij}$ is the separation distance between $i$-th and $j$-th droplets, and $\langle \cdot \rangle$ denotes the ensemble average.

\section*{Acknowledgements}
L.N.C. acknowledges the support of the Postdoctoral EMBO Fellowship ALTF 353-2023 and the TÜBİTAK 2232/B program (project no. 123C289).
For the purpose of open access, the author has applied a Creative Commons Attribution (CC BY) licence to any Author Accepted Manuscript version arising from this submission. 

\section*{Author contributions}
A.U.A., Y.S., G.N. and L.N.C. performed research; A.U.A. and Y.S.
analyzed data;  
G.N. and L.N.C. designed the study;    G.N. and  L.N.C. wrote the paper.

\bibliography{biblio.bib}

\begin{thebibliography}{70}%
\makeatletter
\providecommand \@ifxundefined [1]{%
 \@ifx{#1\undefined}
}%
\providecommand \@ifnum [1]{%
 \ifnum #1\expandafter \@firstoftwo
 \else \expandafter \@secondoftwo
 \fi
}%
\providecommand \@ifx [1]{%
 \ifx #1\expandafter \@firstoftwo
 \else \expandafter \@secondoftwo
 \fi
}%
\providecommand \natexlab [1]{#1}%
\providecommand \enquote  [1]{``#1''}%
\providecommand \bibnamefont  [1]{#1}%
\providecommand \bibfnamefont [1]{#1}%
\providecommand \citenamefont [1]{#1}%
\providecommand \href@noop [0]{\@secondoftwo}%
\providecommand \href [0]{\begingroup \@sanitize@url \@href}%
\providecommand \@href[1]{\@@startlink{#1}\@@href}%
\providecommand \@@href[1]{\endgroup#1\@@endlink}%
\providecommand \@sanitize@url [0]{\catcode `\\12\catcode `\$12\catcode
  `\&12\catcode `\#12\catcode `\^12\catcode `\_12\catcode `\%12\relax}%
\providecommand \@@startlink[1]{}%
\providecommand \@@endlink[0]{}%
\providecommand \url  [0]{\begingroup\@sanitize@url \@url }%
\providecommand \@url [1]{\endgroup\@href {#1}{\urlprefix }}%
\providecommand \urlprefix  [0]{URL }%
\providecommand \Eprint [0]{\href }%
\providecommand \doibase [0]{https://doi.org/}%
\providecommand \selectlanguage [0]{\@gobble}%
\providecommand \bibinfo  [0]{\@secondoftwo}%
\providecommand \bibfield  [0]{\@secondoftwo}%
\providecommand \translation [1]{[#1]}%
\providecommand \BibitemOpen [0]{}%
\providecommand \bibitemStop [0]{}%
\providecommand \bibitemNoStop [0]{.\EOS\space}%
\providecommand \EOS [0]{\spacefactor3000\relax}%
\providecommand \BibitemShut  [1]{\csname bibitem#1\endcsname}%
\let\auto@bib@innerbib\@empty
\bibitem [{\citenamefont {Marchetti}\ \emph {et~al.}(2013)\citenamefont
  {Marchetti}, \citenamefont {Joanny}, \citenamefont {Ramaswamy}, \citenamefont
  {Liverpool}, \citenamefont {Prost}, \citenamefont {Rao},\ and\ \citenamefont
  {Simha}}]{marchetti}%
  \BibitemOpen
  \bibfield  {author} {\bibinfo {author} {\bibfnamefont {M.}~\bibnamefont
  {Marchetti}}, \bibinfo {author} {\bibfnamefont {J.}~\bibnamefont {Joanny}},
  \bibinfo {author} {\bibfnamefont {S.}~\bibnamefont {Ramaswamy}}, \bibinfo
  {author} {\bibfnamefont {T.}~\bibnamefont {Liverpool}}, \bibinfo {author}
  {\bibfnamefont {J.}~\bibnamefont {Prost}}, \bibinfo {author} {\bibfnamefont
  {M.}~\bibnamefont {Rao}},\ and\ \bibinfo {author} {\bibfnamefont
  {R.}~\bibnamefont {Simha}},\ }\bibfield  {title} {\bibinfo {title}
  {Hydrodynamics of soft active matter},\ }\href@noop {} {\bibfield  {journal}
  {\bibinfo  {journal} {Rev. Mod. Phys.}\ }\textbf {\bibinfo {volume} {85}},\
  \bibinfo {pages} {1143} (\bibinfo {year} {2013})}\BibitemShut {NoStop}%
\bibitem [{\citenamefont {Sánchez}\ \emph {et~al.}(2012)\citenamefont
  {Sánchez}, \citenamefont {Chen}, \citenamefont {DeCamp}, \citenamefont
  {Heymann},\ and\ \citenamefont {Dogic}}]{sanchez2012}%
  \BibitemOpen
  \bibfield  {author} {\bibinfo {author} {\bibfnamefont {T.}~\bibnamefont
  {Sánchez}}, \bibinfo {author} {\bibfnamefont {D.~T.~N.}\ \bibnamefont
  {Chen}}, \bibinfo {author} {\bibfnamefont {S.~J.}\ \bibnamefont {DeCamp}},
  \bibinfo {author} {\bibfnamefont {M.}~\bibnamefont {Heymann}},\ and\ \bibinfo
  {author} {\bibfnamefont {Z.}~\bibnamefont {Dogic}},\ }\bibfield  {title}
  {\bibinfo {title} {Spontaneous motion in hierarchically assembled active
  matter},\ }\href {https://doi.org/10.1038/nature11591} {\bibfield  {journal}
  {\bibinfo  {journal} {Nature}\ }\textbf {\bibinfo {volume} {491}},\ \bibinfo
  {pages} {431} (\bibinfo {year} {2012})}\BibitemShut {NoStop}%
\bibitem [{\citenamefont {Zhang}\ \emph {et~al.}(2021)\citenamefont {Zhang},
  \citenamefont {Mozaffari},\ and\ \citenamefont
  {de~Pablo}}]{review_autonomous_materials}%
  \BibitemOpen
  \bibfield  {author} {\bibinfo {author} {\bibfnamefont {R.}~\bibnamefont
  {Zhang}}, \bibinfo {author} {\bibfnamefont {A.}~\bibnamefont {Mozaffari}},\
  and\ \bibinfo {author} {\bibfnamefont {J.~J.}\ \bibnamefont {de~Pablo}},\
  }\bibfield  {title} {\bibinfo {title} {Autonomous materials systems from
  active liquid crystals},\ }\href {https://doi.org/10.1038/s41578-021-00305-z}
  {\bibfield  {journal} {\bibinfo  {journal} {Nat. Rev. Mater.}\ }\textbf
  {\bibinfo {volume} {6}},\ \bibinfo {pages} {437} (\bibinfo {year}
  {2021})}\BibitemShut {NoStop}%
\bibitem [{\citenamefont {Doostmohammadi}\ \emph {et~al.}(2018)\citenamefont
  {Doostmohammadi}, \citenamefont {Ignés-Mullol}, \citenamefont {Yeomans},\
  and\ \citenamefont {Sagués}}]{doostmohammadi2018}%
  \BibitemOpen
  \bibfield  {author} {\bibinfo {author} {\bibfnamefont {A.}~\bibnamefont
  {Doostmohammadi}}, \bibinfo {author} {\bibfnamefont {J.}~\bibnamefont
  {Ignés-Mullol}}, \bibinfo {author} {\bibfnamefont {J.~M.}\ \bibnamefont
  {Yeomans}},\ and\ \bibinfo {author} {\bibfnamefont {F.}~\bibnamefont
  {Sagués}},\ }\bibfield  {title} {\bibinfo {title} {Active nematics},\ }\href
  {https://doi.org/10.1038/s41467-018-05666-8} {\bibfield  {journal} {\bibinfo
  {journal} {Nat. Commun.}\ }\textbf {\bibinfo {volume} {9}},\ \bibinfo {pages}
  {3246} (\bibinfo {year} {2018})}\BibitemShut {NoStop}%
\bibitem [{\citenamefont {Voituriez}\ \emph {et~al.}(2006)\citenamefont
  {Voituriez}, \citenamefont {Joanny},\ and\ \citenamefont
  {Prost}}]{Voituriez2006}%
  \BibitemOpen
  \bibfield  {author} {\bibinfo {author} {\bibfnamefont {R.}~\bibnamefont
  {Voituriez}}, \bibinfo {author} {\bibfnamefont {J.~F.}\ \bibnamefont
  {Joanny}},\ and\ \bibinfo {author} {\bibfnamefont {J.}~\bibnamefont
  {Prost}},\ }\bibfield  {title} {\bibinfo {title} {Generic phase diagram of
  active polar films},\ }\href {https://doi.org/10.1103/PhysRevLett.96.028102}
  {\bibfield  {journal} {\bibinfo  {journal} {Phys. Rev. Lett.}\ }\textbf
  {\bibinfo {volume} {96}},\ \bibinfo {pages} {028102} (\bibinfo {year}
  {2006})}\BibitemShut {NoStop}%
\bibitem [{\citenamefont {Giomi}\ \emph {et~al.}(2011)\citenamefont {Giomi},
  \citenamefont {Mahadevan}, \citenamefont {Chakraborty},\ and\ \citenamefont
  {Hagan}}]{giomi2011}%
  \BibitemOpen
  \bibfield  {author} {\bibinfo {author} {\bibfnamefont {L.}~\bibnamefont
  {Giomi}}, \bibinfo {author} {\bibfnamefont {L.}~\bibnamefont {Mahadevan}},
  \bibinfo {author} {\bibfnamefont {B.}~\bibnamefont {Chakraborty}},\ and\
  \bibinfo {author} {\bibfnamefont {M.~F.}\ \bibnamefont {Hagan}},\ }\bibfield
  {title} {\bibinfo {title} {Excitable patterns in active nematics},\ }\href
  {https://doi.org/10.1103/PhysRevLett.106.218101} {\bibfield  {journal}
  {\bibinfo  {journal} {Phys. Rev. Lett.}\ }\textbf {\bibinfo {volume} {106}},\
  \bibinfo {pages} {218101} (\bibinfo {year} {2011})}\BibitemShut {NoStop}%
\bibitem [{\citenamefont {Thampi}\ \emph {et~al.}(2013)\citenamefont {Thampi},
  \citenamefont {Golestanian},\ and\ \citenamefont {Yeomans}}]{Thampi2013}%
  \BibitemOpen
  \bibfield  {author} {\bibinfo {author} {\bibfnamefont {S.~P.}\ \bibnamefont
  {Thampi}}, \bibinfo {author} {\bibfnamefont {R.}~\bibnamefont
  {Golestanian}},\ and\ \bibinfo {author} {\bibfnamefont {J.~M.}\ \bibnamefont
  {Yeomans}},\ }\bibfield  {title} {\bibinfo {title} {Velocity correlations in
  an active nematic},\ }\href {https://doi.org/10.1103/PhysRevLett.111.118101}
  {\bibfield  {journal} {\bibinfo  {journal} {Phys. Rev. Lett.}\ }\textbf
  {\bibinfo {volume} {111}},\ \bibinfo {pages} {118101} (\bibinfo {year}
  {2013})}\BibitemShut {NoStop}%
\bibitem [{\citenamefont {Giomi}\ \emph {et~al.}(2013)\citenamefont {Giomi},
  \citenamefont {Bowick}, \citenamefont {Ma},\ and\ \citenamefont
  {Marchetti}}]{giomi2013}%
  \BibitemOpen
  \bibfield  {author} {\bibinfo {author} {\bibfnamefont {L.}~\bibnamefont
  {Giomi}}, \bibinfo {author} {\bibfnamefont {M.~J.}\ \bibnamefont {Bowick}},
  \bibinfo {author} {\bibfnamefont {X.}~\bibnamefont {Ma}},\ and\ \bibinfo
  {author} {\bibfnamefont {M.~C.}\ \bibnamefont {Marchetti}},\ }\bibfield
  {title} {\bibinfo {title} {Defect annihilation and proliferation in active
  nematics},\ }\href {https://doi.org/10.1103/PhysRevLett.110.228101}
  {\bibfield  {journal} {\bibinfo  {journal} {Phys. Rev. Lett.}\ }\textbf
  {\bibinfo {volume} {110}},\ \bibinfo {pages} {228101} (\bibinfo {year}
  {2013})}\BibitemShut {NoStop}%
\bibitem [{\citenamefont {Ngo}\ \emph {et~al.}(2014)\citenamefont {Ngo},
  \citenamefont {Peshkov}, \citenamefont {Aranson}, \citenamefont {Bertin},
  \citenamefont {Ginelli},\ and\ \citenamefont {Chat\'e}}]{Ngo2014}%
  \BibitemOpen
  \bibfield  {author} {\bibinfo {author} {\bibfnamefont {S.}~\bibnamefont
  {Ngo}}, \bibinfo {author} {\bibfnamefont {A.}~\bibnamefont {Peshkov}},
  \bibinfo {author} {\bibfnamefont {I.~S.}\ \bibnamefont {Aranson}}, \bibinfo
  {author} {\bibfnamefont {E.}~\bibnamefont {Bertin}}, \bibinfo {author}
  {\bibfnamefont {F.}~\bibnamefont {Ginelli}},\ and\ \bibinfo {author}
  {\bibfnamefont {H.}~\bibnamefont {Chat\'e}},\ }\bibfield  {title} {\bibinfo
  {title} {Large-scale chaos and fluctuations in active nematics},\ }\href
  {https://doi.org/10.1103/PhysRevLett.113.038302} {\bibfield  {journal}
  {\bibinfo  {journal} {Phys. Rev. Lett.}\ }\textbf {\bibinfo {volume} {113}},\
  \bibinfo {pages} {038302} (\bibinfo {year} {2014})}\BibitemShut {NoStop}%
\bibitem [{\citenamefont {Thampi}\ \emph {et~al.}(2014)\citenamefont {Thampi},
  \citenamefont {Golestanian},\ and\ \citenamefont {Yeomans}}]{Thampi_2014}%
  \BibitemOpen
  \bibfield  {author} {\bibinfo {author} {\bibfnamefont {S.~P.}\ \bibnamefont
  {Thampi}}, \bibinfo {author} {\bibfnamefont {R.}~\bibnamefont
  {Golestanian}},\ and\ \bibinfo {author} {\bibfnamefont {J.~M.}\ \bibnamefont
  {Yeomans}},\ }\bibfield  {title} {\bibinfo {title} {Instabilities and
  topological defects in active nematics},\ }\href
  {https://doi.org/10.1209/0295-5075/105/18001} {\bibfield  {journal} {\bibinfo
   {journal} {EPL}\ }\textbf {\bibinfo {volume} {105}},\ \bibinfo {pages}
  {18001} (\bibinfo {year} {2014})}\BibitemShut {NoStop}%
\bibitem [{\citenamefont {Shankar}\ \emph {et~al.}(2018)\citenamefont
  {Shankar}, \citenamefont {Ramaswamy}, \citenamefont {Marchetti},\ and\
  \citenamefont {Bowick}}]{Shankar2018}%
  \BibitemOpen
  \bibfield  {author} {\bibinfo {author} {\bibfnamefont {S.}~\bibnamefont
  {Shankar}}, \bibinfo {author} {\bibfnamefont {S.}~\bibnamefont {Ramaswamy}},
  \bibinfo {author} {\bibfnamefont {M.~C.}\ \bibnamefont {Marchetti}},\ and\
  \bibinfo {author} {\bibfnamefont {M.~J.}\ \bibnamefont {Bowick}},\ }\bibfield
   {title} {\bibinfo {title} {Defect unbinding in active nematics},\ }\href
  {https://doi.org/10.1103/PhysRevLett.121.108002} {\bibfield  {journal}
  {\bibinfo  {journal} {Phys. Rev. Lett.}\ }\textbf {\bibinfo {volume} {121}},\
  \bibinfo {pages} {108002} (\bibinfo {year} {2018})}\BibitemShut {NoStop}%
\bibitem [{\citenamefont {Alert}\ \emph {et~al.}(2022)\citenamefont {Alert},
  \citenamefont {Casademunt},\ and\ \citenamefont {Joanny}}]{Alert2022}%
  \BibitemOpen
  \bibfield  {author} {\bibinfo {author} {\bibfnamefont {R.}~\bibnamefont
  {Alert}}, \bibinfo {author} {\bibfnamefont {J.}~\bibnamefont {Casademunt}},\
  and\ \bibinfo {author} {\bibfnamefont {J.-F.}\ \bibnamefont {Joanny}},\
  }\bibfield  {title} {\bibinfo {title} {Active turbulence},\ }\href
  {https://doi.org/10.1146/annurev-conmatphys-082321-035957} {\bibfield
  {journal} {\bibinfo  {journal} {Annu. Rev. Condens. Matter Phys.}\ }\textbf
  {\bibinfo {volume} {13}},\ \bibinfo {pages} {143} (\bibinfo {year}
  {2022})}\BibitemShut {NoStop}%
\bibitem [{\citenamefont {Head}\ \emph
  {et~al.}(2024{\natexlab{a}})\citenamefont {Head}, \citenamefont {Doré},
  \citenamefont {Keogh}, \citenamefont {Bonn}, \citenamefont {Negro},
  \citenamefont {Marenduzzo}, \citenamefont {Doostmohammadi}, \citenamefont
  {Thijssen}, \citenamefont {López-León},\ and\ \citenamefont
  {Shendruk}}]{Head2024}%
  \BibitemOpen
  \bibfield  {author} {\bibinfo {author} {\bibfnamefont {L.~C.}\ \bibnamefont
  {Head}}, \bibinfo {author} {\bibfnamefont {C.}~\bibnamefont {Doré}},
  \bibinfo {author} {\bibfnamefont {R.~R.}\ \bibnamefont {Keogh}}, \bibinfo
  {author} {\bibfnamefont {L.}~\bibnamefont {Bonn}}, \bibinfo {author}
  {\bibfnamefont {G.}~\bibnamefont {Negro}}, \bibinfo {author} {\bibfnamefont
  {D.}~\bibnamefont {Marenduzzo}}, \bibinfo {author} {\bibfnamefont
  {A.}~\bibnamefont {Doostmohammadi}}, \bibinfo {author} {\bibfnamefont
  {K.}~\bibnamefont {Thijssen}}, \bibinfo {author} {\bibfnamefont
  {T.}~\bibnamefont {López-León}},\ and\ \bibinfo {author} {\bibfnamefont
  {T.~N.}\ \bibnamefont {Shendruk}},\ }\bibfield  {title} {\bibinfo {title}
  {Spontaneous self-constraint in active nematic flows},\ }\href
  {https://doi.org/10.1038/s41567-023-02336-5} {\bibfield  {journal} {\bibinfo
  {journal} {Nat. Phys.}\ }\textbf {\bibinfo {volume} {20}},\ \bibinfo {pages}
  {231} (\bibinfo {year} {2024}{\natexlab{a}})}\BibitemShut {NoStop}%
\bibitem [{\citenamefont {Carenza}\ \emph {et~al.}(2025)\citenamefont
  {Carenza}, \citenamefont {Armengol-Collado}, \citenamefont {Krommydas},\ and\
  \citenamefont {Giomi}}]{carenza2025}%
  \BibitemOpen
  \bibfield  {author} {\bibinfo {author} {\bibfnamefont {L.~N.}\ \bibnamefont
  {Carenza}}, \bibinfo {author} {\bibfnamefont {J.-M.}\ \bibnamefont
  {Armengol-Collado}}, \bibinfo {author} {\bibfnamefont {D.}~\bibnamefont
  {Krommydas}},\ and\ \bibinfo {author} {\bibfnamefont {L.}~\bibnamefont
  {Giomi}},\ }\bibfield  {title} {\bibinfo {title} {Quasi-long-ranged order in
  two-dimensional active liquid crystals},\ }\href
  {https://doi.org/10.1103/PhysRevLett.134.128304} {\bibfield  {journal}
  {\bibinfo  {journal} {Phys. Rev. Lett.}\ }\textbf {\bibinfo {volume} {134}},\
  \bibinfo {pages} {128304} (\bibinfo {year} {2025})}\BibitemShut {NoStop}%
\bibitem [{\citenamefont {Ray}\ \emph {et~al.}(2023)\citenamefont {Ray},
  \citenamefont {Zhang},\ and\ \citenamefont {Dogic}}]{Ray2023}%
  \BibitemOpen
  \bibfield  {author} {\bibinfo {author} {\bibfnamefont {S.}~\bibnamefont
  {Ray}}, \bibinfo {author} {\bibfnamefont {J.}~\bibnamefont {Zhang}},\ and\
  \bibinfo {author} {\bibfnamefont {Z.}~\bibnamefont {Dogic}},\ }\bibfield
  {title} {\bibinfo {title} {Rectified rotational dynamics of mobile inclusions
  in two-dimensional active nematics},\ }\href
  {https://doi.org/10.1103/PhysRevLett.130.238301} {\bibfield  {journal}
  {\bibinfo  {journal} {Phys. Rev. Lett.}\ }\textbf {\bibinfo {volume} {130}},\
  \bibinfo {pages} {238301} (\bibinfo {year} {2023})}\BibitemShut {NoStop}%
\bibitem [{\citenamefont {Neville}\ \emph {et~al.}(2024)\citenamefont
  {Neville}, \citenamefont {Eggers},\ and\ \citenamefont
  {Liverpool}}]{Neville2024}%
  \BibitemOpen
  \bibfield  {author} {\bibinfo {author} {\bibfnamefont {L.}~\bibnamefont
  {Neville}}, \bibinfo {author} {\bibfnamefont {J.}~\bibnamefont {Eggers}},\
  and\ \bibinfo {author} {\bibfnamefont {T.~B.}\ \bibnamefont {Liverpool}},\
  }\bibfield  {title} {\bibinfo {title} {Controlling wall–particle
  interactions with activity},\ }\href {https://doi.org/10.1039/D4SM00634H}
  {\bibfield  {journal} {\bibinfo  {journal} {Soft Matter}\ }\textbf {\bibinfo
  {volume} {20}},\ \bibinfo {pages} {8395} (\bibinfo {year}
  {2024})}\BibitemShut {NoStop}%
\bibitem [{\citenamefont {Loewe}\ and\ \citenamefont
  {Shendruk}(2022)}]{Loewe_2022}%
  \BibitemOpen
  \bibfield  {author} {\bibinfo {author} {\bibfnamefont {B.}~\bibnamefont
  {Loewe}}\ and\ \bibinfo {author} {\bibfnamefont {T.~N.}\ \bibnamefont
  {Shendruk}},\ }\bibfield  {title} {\bibinfo {title} {Passive {Janus}
  particles are self-propelled in active nematics},\ }\href
  {https://doi.org/10.1088/1367-2630/ac3b70} {\bibfield  {journal} {\bibinfo
  {journal} {New J. Phys.}\ }\textbf {\bibinfo {volume} {24}},\ \bibinfo
  {pages} {012001} (\bibinfo {year} {2022})}\BibitemShut {NoStop}%
\bibitem [{\citenamefont {Chandler}\ and\ \citenamefont
  {Spagnolie}(2024{\natexlab{a}})}]{chandler2024_2}%
  \BibitemOpen
  \bibfield  {author} {\bibinfo {author} {\bibfnamefont {T.~G.~J.}\
  \bibnamefont {Chandler}}\ and\ \bibinfo {author} {\bibfnamefont {S.~E.}\
  \bibnamefont {Spagnolie}},\ }\bibfield  {title} {\bibinfo {title} {Active
  nematic response to a deformable body or boundary: Elastic deformations and
  anchoring-induced flow},\ }\bibfield  {journal} {\bibinfo  {journal} {arXiv
  preprint}\ }\href {https://doi.org/10.48550/arXiv.2409.15617}
  {10.48550/arXiv.2409.15617} (\bibinfo {year} {2024}{\natexlab{a}}),\ \Eprint
  {https://arxiv.org/abs/2409.15617} {arXiv:2409.15617 [cond-mat.soft]}
  \BibitemShut {NoStop}%
\bibitem [{\citenamefont {Thampi}\ \emph {et~al.}(2016)\citenamefont {Thampi},
  \citenamefont {Doostmohammadi}, \citenamefont {Shendruk}, \citenamefont
  {Golestanian},\ and\ \citenamefont {Yeomans}}]{Thampi2016}%
  \BibitemOpen
  \bibfield  {author} {\bibinfo {author} {\bibfnamefont {S.~P.}\ \bibnamefont
  {Thampi}}, \bibinfo {author} {\bibfnamefont {A.}~\bibnamefont
  {Doostmohammadi}}, \bibinfo {author} {\bibfnamefont {T.~N.}\ \bibnamefont
  {Shendruk}}, \bibinfo {author} {\bibfnamefont {R.}~\bibnamefont
  {Golestanian}},\ and\ \bibinfo {author} {\bibfnamefont {J.~M.}\ \bibnamefont
  {Yeomans}},\ }\bibfield  {title} {\bibinfo {title} {Active micromachines:
  Microfluidics powered by mesoscale turbulence},\ }\href
  {https://doi.org/10.1126/sciadv.1501854} {\bibfield  {journal} {\bibinfo
  {journal} {Sci. Adv.}\ }\textbf {\bibinfo {volume} {2}},\ \bibinfo {pages}
  {e1501854} (\bibinfo {year} {2016})}\BibitemShut {NoStop}%
\bibitem [{\citenamefont {Nishiguchi}\ \emph {et~al.}(2018)\citenamefont
  {Nishiguchi}, \citenamefont {Aranson}, \citenamefont {Snezhko},\ and\
  \citenamefont {Sokolov}}]{Nishiguchi2018}%
  \BibitemOpen
  \bibfield  {author} {\bibinfo {author} {\bibfnamefont {D.}~\bibnamefont
  {Nishiguchi}}, \bibinfo {author} {\bibfnamefont {I.~S.}\ \bibnamefont
  {Aranson}}, \bibinfo {author} {\bibfnamefont {A.}~\bibnamefont {Snezhko}},\
  and\ \bibinfo {author} {\bibfnamefont {A.}~\bibnamefont {Sokolov}},\
  }\bibfield  {title} {\bibinfo {title} {Engineering bacterial vortex lattice
  via direct laser lithography},\ }\href
  {https://doi.org/10.1038/s41467-018-06842-6} {\bibfield  {journal} {\bibinfo
  {journal} {Nat. Commun.}\ }\textbf {\bibinfo {volume} {9}},\ \bibinfo {pages}
  {4486} (\bibinfo {year} {2018})}\BibitemShut {NoStop}%
\bibitem [{\citenamefont {Di~Leonardo}\ \emph {et~al.}(2009)\citenamefont
  {Di~Leonardo}, \citenamefont {Angelani}, \citenamefont {Dell'Arciprete},
  \citenamefont {Ruocco}, \citenamefont {Iebba}, \citenamefont {Schippa},
  \citenamefont {Conte}, \citenamefont {Mecarini}, \citenamefont {De~Angelis},\
  and\ \citenamefont {Di~Fabrizio}}]{Leonardo2009}%
  \BibitemOpen
  \bibfield  {author} {\bibinfo {author} {\bibfnamefont {R.}~\bibnamefont
  {Di~Leonardo}}, \bibinfo {author} {\bibfnamefont {L.}~\bibnamefont
  {Angelani}}, \bibinfo {author} {\bibfnamefont {D.}~\bibnamefont
  {Dell'Arciprete}}, \bibinfo {author} {\bibfnamefont {G.}~\bibnamefont
  {Ruocco}}, \bibinfo {author} {\bibfnamefont {V.}~\bibnamefont {Iebba}},
  \bibinfo {author} {\bibfnamefont {S.}~\bibnamefont {Schippa}}, \bibinfo
  {author} {\bibfnamefont {M.~P.}\ \bibnamefont {Conte}}, \bibinfo {author}
  {\bibfnamefont {F.}~\bibnamefont {Mecarini}}, \bibinfo {author}
  {\bibfnamefont {F.}~\bibnamefont {De~Angelis}},\ and\ \bibinfo {author}
  {\bibfnamefont {E.}~\bibnamefont {Di~Fabrizio}},\ }\bibfield  {title}
  {\bibinfo {title} {Bacterial ratchet motors},\ }\href
  {https://doi.org/10.1073/pnas.0910426107} {\bibfield  {journal} {\bibinfo
  {journal} {Proc. Natl. Acad. Sci. U.S.A.}\ }\textbf {\bibinfo {volume}
  {107}},\ \bibinfo {pages} {9541} (\bibinfo {year} {2009})}\BibitemShut
  {NoStop}%
\bibitem [{\citenamefont {Reinken}\ \emph {et~al.}(2020)\citenamefont
  {Reinken}, \citenamefont {Nishiguchi}, \citenamefont {Heidenreich},
  \citenamefont {Sokolov}, \citenamefont {Bär}, \citenamefont {Klapp},\ and\
  \citenamefont {Aranson}}]{Reinken2020}%
  \BibitemOpen
  \bibfield  {author} {\bibinfo {author} {\bibfnamefont {H.}~\bibnamefont
  {Reinken}}, \bibinfo {author} {\bibfnamefont {D.}~\bibnamefont {Nishiguchi}},
  \bibinfo {author} {\bibfnamefont {S.}~\bibnamefont {Heidenreich}}, \bibinfo
  {author} {\bibfnamefont {A.}~\bibnamefont {Sokolov}}, \bibinfo {author}
  {\bibfnamefont {M.}~\bibnamefont {Bär}}, \bibinfo {author} {\bibfnamefont
  {S.~H.~L.}\ \bibnamefont {Klapp}},\ and\ \bibinfo {author} {\bibfnamefont
  {I.~S.}\ \bibnamefont {Aranson}},\ }\bibfield  {title} {\bibinfo {title}
  {Organizing bacterial vortex lattices by periodic obstacle arrays},\ }\href
  {https://doi.org/10.1038/s42005-020-0337-z} {\bibfield  {journal} {\bibinfo
  {journal} {Commun. Phys.}\ }\textbf {\bibinfo {volume} {3}},\ \bibinfo
  {pages} {76} (\bibinfo {year} {2020})}\BibitemShut {NoStop}%
\bibitem [{\citenamefont {{Carenza, L. N.}}\ \emph {et~al.}(2020)\citenamefont
  {{Carenza, L. N.}}, \citenamefont {{Biferale, L.}},\ and\ \citenamefont
  {{Gonnella, G.}}}]{carenzaEPL_2020}%
  \BibitemOpen
  \bibfield  {author} {\bibinfo {author} {\bibnamefont {{Carenza, L. N.}}},
  \bibinfo {author} {\bibnamefont {{Biferale, L.}}},\ and\ \bibinfo {author}
  {\bibnamefont {{Gonnella, G.}}},\ }\bibfield  {title} {\bibinfo {title}
  {Cascade or not cascade? energy transfer and elastic effects in active
  nematics},\ }\href {https://doi.org/10.1209/0295-5075/132/44003} {\bibfield
  {journal} {\bibinfo  {journal} {EPL}\ }\textbf {\bibinfo {volume} {132}},\
  \bibinfo {pages} {44003} (\bibinfo {year} {2020})}\BibitemShut {NoStop}%
\bibitem [{\citenamefont {Houston}\ and\ \citenamefont
  {Alexander}(2023)}]{Houston_2023}%
  \BibitemOpen
  \bibfield  {author} {\bibinfo {author} {\bibfnamefont {A.~J.~H.}\
  \bibnamefont {Houston}}\ and\ \bibinfo {author} {\bibfnamefont {G.~P.}\
  \bibnamefont {Alexander}},\ }\bibfield  {title} {\bibinfo {title} {Colloids
  in two-dimensional active nematics: conformal cogs and controllable
  spontaneous rotation},\ }\href {https://doi.org/10.1088/1367-2630/ad0acf}
  {\bibfield  {journal} {\bibinfo  {journal} {New J. Phys.}\ }\textbf {\bibinfo
  {volume} {25}},\ \bibinfo {pages} {123006} (\bibinfo {year}
  {2023})}\BibitemShut {NoStop}%
\bibitem [{\citenamefont {Vélez-Ceron}\ \emph {et~al.}(2024)\citenamefont
  {Vélez-Ceron}, \citenamefont {Coelho}, \citenamefont {Guillamat},
  \citenamefont {da~Gama}, \citenamefont {Sagués},\ and\ \citenamefont
  {Ignés-Mullol}}]{VelezCeron2024}%
  \BibitemOpen
  \bibfield  {author} {\bibinfo {author} {\bibfnamefont {I.}~\bibnamefont
  {Vélez-Ceron}}, \bibinfo {author} {\bibfnamefont {R.~C.~V.}\ \bibnamefont
  {Coelho}}, \bibinfo {author} {\bibfnamefont {P.}~\bibnamefont {Guillamat}},
  \bibinfo {author} {\bibfnamefont {M.~T.}\ \bibnamefont {da~Gama}}, \bibinfo
  {author} {\bibfnamefont {F.}~\bibnamefont {Sagués}},\ and\ \bibinfo {author}
  {\bibfnamefont {J.}~\bibnamefont {Ignés-Mullol}},\ }\bibfield  {title}
  {\bibinfo {title} {Active nematic pumps},\ }\href
  {https://arxiv.org/abs/2407.09960} {\bibfield  {journal} {\bibinfo  {journal}
  {arXiv preprint}\ } (\bibinfo {year} {2024})},\ \Eprint
  {https://arxiv.org/abs/2407.09960} {arXiv:2407.09960 [cond-mat.soft]}
  \BibitemShut {NoStop}%
\bibitem [{\citenamefont {Kozhukhov}\ \emph {et~al.}(2025)\citenamefont
  {Kozhukhov}, \citenamefont {Loewe}, \citenamefont {Thijssen},\ and\
  \citenamefont {Shendruk}}]{kozhukhov2025}%
  \BibitemOpen
  \bibfield  {author} {\bibinfo {author} {\bibfnamefont {T.}~\bibnamefont
  {Kozhukhov}}, \bibinfo {author} {\bibfnamefont {B.}~\bibnamefont {Loewe}},
  \bibinfo {author} {\bibfnamefont {K.}~\bibnamefont {Thijssen}},\ and\
  \bibinfo {author} {\bibfnamefont {T.~N.}\ \bibnamefont {Shendruk}},\
  }\bibfield  {title} {\bibinfo {title} {Topological kicks enhance colloidal
  diffusivity in topological turbulence},\ }\href
  {https://arxiv.org/abs/2503.19664} {\bibfield  {journal} {\bibinfo  {journal}
  {arXiv preprint}\ } (\bibinfo {year} {2025})},\ \Eprint
  {https://arxiv.org/abs/2503.19664} {arXiv:2503.19664 [cond-mat.soft]}
  \BibitemShut {NoStop}%
\bibitem [{\citenamefont {Schimming}\ \emph {et~al.}(2024)\citenamefont
  {Schimming}, \citenamefont {Reichhardt},\ and\ \citenamefont
  {Reichhardt}}]{Schimming2024}%
  \BibitemOpen
  \bibfield  {author} {\bibinfo {author} {\bibfnamefont {C.~D.}\ \bibnamefont
  {Schimming}}, \bibinfo {author} {\bibfnamefont {C.~J.~O.}\ \bibnamefont
  {Reichhardt}},\ and\ \bibinfo {author} {\bibfnamefont {C.}~\bibnamefont
  {Reichhardt}},\ }\bibfield  {title} {\bibinfo {title} {Active nematic ratchet
  in asymmetric obstacle arrays},\ }\href
  {https://doi.org/10.1103/PhysRevE.109.064602} {\bibfield  {journal} {\bibinfo
   {journal} {Phys. Rev. E}\ }\textbf {\bibinfo {volume} {109}},\ \bibinfo
  {pages} {064602} (\bibinfo {year} {2024})}\BibitemShut {NoStop}%
\bibitem [{\citenamefont {Carenza}\ \emph
  {et~al.}(2020{\natexlab{a}})\citenamefont {Carenza}, \citenamefont
  {Biferale},\ and\ \citenamefont {Gonnella}}]{Carenza2020}%
  \BibitemOpen
  \bibfield  {author} {\bibinfo {author} {\bibfnamefont {L.}~\bibnamefont
  {Carenza}}, \bibinfo {author} {\bibfnamefont {L.}~\bibnamefont {Biferale}},\
  and\ \bibinfo {author} {\bibfnamefont {G.}~\bibnamefont {Gonnella}},\
  }\bibfield  {title} {\bibinfo {title} {Multiscale control of active emulsion
  dynamics},\ }\href {https://doi.org/10.1103/PhysRevFluids.5.011302}
  {\bibfield  {journal} {\bibinfo  {journal} {Phys. Rev. Fluids}\ }\textbf
  {\bibinfo {volume} {5}},\ \bibinfo {pages} {11302} (\bibinfo {year}
  {2020}{\natexlab{a}})}\BibitemShut {NoStop}%
\bibitem [{\citenamefont {Stark}(2001)}]{Stark2001}%
  \BibitemOpen
  \bibfield  {author} {\bibinfo {author} {\bibfnamefont {H.}~\bibnamefont
  {Stark}},\ }\bibfield  {title} {\bibinfo {title} {Physics of colloidal
  dispersions in nematic liquid crystals},\ }\href
  {https://doi.org/https://doi.org/10.1016/S0370-1573(00)00144-7} {\bibfield
  {journal} {\bibinfo  {journal} {Phys. Rep.}\ }\textbf {\bibinfo {volume}
  {351}},\ \bibinfo {pages} {387} (\bibinfo {year} {2001})}\BibitemShut
  {NoStop}%
\bibitem [{\citenamefont {Smalyukh}\ \emph {et~al.}(2005)\citenamefont
  {Smalyukh}, \citenamefont {Lavrentovich}, \citenamefont {Kuzmin},
  \citenamefont {Kachynski},\ and\ \citenamefont {Prasad}}]{Smalyuk2005}%
  \BibitemOpen
  \bibfield  {author} {\bibinfo {author} {\bibfnamefont {I.~I.}\ \bibnamefont
  {Smalyukh}}, \bibinfo {author} {\bibfnamefont {O.~D.}\ \bibnamefont
  {Lavrentovich}}, \bibinfo {author} {\bibfnamefont {A.~N.}\ \bibnamefont
  {Kuzmin}}, \bibinfo {author} {\bibfnamefont {A.~V.}\ \bibnamefont
  {Kachynski}},\ and\ \bibinfo {author} {\bibfnamefont {P.~N.}\ \bibnamefont
  {Prasad}},\ }\bibfield  {title} {\bibinfo {title} {Elasticity-mediated
  self-organization and colloidal interactions of solid spheres with tangential
  anchoring in a nematic liquid crystal},\ }\href
  {https://doi.org/10.1103/PhysRevLett.95.157801} {\bibfield  {journal}
  {\bibinfo  {journal} {Phys. Rev. Lett.}\ }\textbf {\bibinfo {volume} {95}},\
  \bibinfo {pages} {157801} (\bibinfo {year} {2005})}\BibitemShut {NoStop}%
\bibitem [{\citenamefont {Smalyukh}(2018)}]{Smalyuk2018}%
  \BibitemOpen
  \bibfield  {author} {\bibinfo {author} {\bibfnamefont {I.~I.}\ \bibnamefont
  {Smalyukh}},\ }\bibfield  {title} {\bibinfo {title} {Liquid crystal
  colloids},\ }\href
  {https://doi.org/https://doi.org/10.1146/annurev-conmatphys-033117-054102}
  {\bibfield  {journal} {\bibinfo  {journal} {Annu. Rev. Condens. Matter
  Phys.}\ }\textbf {\bibinfo {volume} {9}},\ \bibinfo {pages} {207} (\bibinfo
  {year} {2018})}\BibitemShut {NoStop}%
\bibitem [{\citenamefont {Genkin}\ \emph {et~al.}(2018)\citenamefont {Genkin},
  \citenamefont {Sokolov},\ and\ \citenamefont {Aranson}}]{Genkin2018}%
  \BibitemOpen
  \bibfield  {author} {\bibinfo {author} {\bibfnamefont {M.~M.}\ \bibnamefont
  {Genkin}}, \bibinfo {author} {\bibfnamefont {A.}~\bibnamefont {Sokolov}},\
  and\ \bibinfo {author} {\bibfnamefont {I.~S.}\ \bibnamefont {Aranson}},\
  }\bibfield  {title} {\bibinfo {title} {Spontaneous topological charging of
  tactoids in a living nematic},\ }\href
  {https://doi.org/10.1088/1367-2630/aab1a3} {\bibfield  {journal} {\bibinfo
  {journal} {New J. Phys.}\ }\textbf {\bibinfo {volume} {20}},\ \bibinfo
  {pages} {043027} (\bibinfo {year} {2018})}\BibitemShut {NoStop}%
\bibitem [{\citenamefont {Kim}\ \emph {et~al.}(2018)\citenamefont {Kim},
  \citenamefont {Wang}, \citenamefont {Mondkar}, \citenamefont {Bukusoglu},\
  and\ \citenamefont {Abbott}}]{Kim2018}%
  \BibitemOpen
  \bibfield  {author} {\bibinfo {author} {\bibfnamefont {Y.-K.}\ \bibnamefont
  {Kim}}, \bibinfo {author} {\bibfnamefont {X.}~\bibnamefont {Wang}}, \bibinfo
  {author} {\bibfnamefont {P.}~\bibnamefont {Mondkar}}, \bibinfo {author}
  {\bibfnamefont {E.}~\bibnamefont {Bukusoglu}},\ and\ \bibinfo {author}
  {\bibfnamefont {N.~L.}\ \bibnamefont {Abbott}},\ }\bibfield  {title}
  {\bibinfo {title} {Self-reporting and self-regulating liquid crystals},\
  }\href {https://doi.org/10.1038/s41586-018-0098-y} {\bibfield  {journal}
  {\bibinfo  {journal} {Nature}\ }\textbf {\bibinfo {volume} {557}},\ \bibinfo
  {pages} {539} (\bibinfo {year} {2018})}\BibitemShut {NoStop}%
\bibitem [{\citenamefont {Senyuk}\ \emph {et~al.}(2022)\citenamefont {Senyuk},
  \citenamefont {Meng},\ and\ \citenamefont {Smalyukh}}]{senyuk2022}%
  \BibitemOpen
  \bibfield  {author} {\bibinfo {author} {\bibfnamefont {B.}~\bibnamefont
  {Senyuk}}, \bibinfo {author} {\bibfnamefont {C.}~\bibnamefont {Meng}},\ and\
  \bibinfo {author} {\bibfnamefont {I.~I.}\ \bibnamefont {Smalyukh}},\
  }\bibfield  {title} {\bibinfo {title} {Design and preparation of nematic
  colloidal particles},\ }\href {https://doi.org/10.1021/acs.langmuir.2c00611}
  {\bibfield  {journal} {\bibinfo  {journal} {Langmuir}\ }\textbf {\bibinfo
  {volume} {38}},\ \bibinfo {pages} {9099} (\bibinfo {year}
  {2022})}\BibitemShut {NoStop}%
\bibitem [{\citenamefont {Muševič}(2013)}]{Musevich2013}%
  \BibitemOpen
  \bibfield  {author} {\bibinfo {author} {\bibfnamefont {I.}~\bibnamefont
  {Muševič}},\ }\bibfield  {title} {\bibinfo {title} {Nematic colloids,
  topology and photonics},\ }\href {https://doi.org/10.1098/rsta.2012.0266}
  {\bibfield  {journal} {\bibinfo  {journal} {Phil. Trans. R. Soc. A}\ }\textbf
  {\bibinfo {volume} {371}},\ \bibinfo {pages} {20120266} (\bibinfo {year}
  {2013})}\BibitemShut {NoStop}%
\bibitem [{\citenamefont {Poulin}\ \emph {et~al.}(1997)\citenamefont {Poulin},
  \citenamefont {Stark}, \citenamefont {Lubensky},\ and\ \citenamefont
  {Weitz}}]{Poulin1997}%
  \BibitemOpen
  \bibfield  {author} {\bibinfo {author} {\bibfnamefont {P.}~\bibnamefont
  {Poulin}}, \bibinfo {author} {\bibfnamefont {H.}~\bibnamefont {Stark}},
  \bibinfo {author} {\bibfnamefont {T.~C.}\ \bibnamefont {Lubensky}},\ and\
  \bibinfo {author} {\bibfnamefont {D.~A.}\ \bibnamefont {Weitz}},\ }\bibfield
  {title} {\bibinfo {title} {Novel colloidal interactions in anisotropic
  fluids},\ }\href {https://doi.org/10.1126/science.275.5307.1770} {\bibfield
  {journal} {\bibinfo  {journal} {Science}\ }\textbf {\bibinfo {volume}
  {275}},\ \bibinfo {pages} {1770} (\bibinfo {year} {1997})}\BibitemShut
  {NoStop}%
\bibitem [{\citenamefont {Lubensky}\ \emph {et~al.}(1998)\citenamefont
  {Lubensky}, \citenamefont {Pettey}, \citenamefont {Currier},\ and\
  \citenamefont {Stark}}]{Lubensky1998}%
  \BibitemOpen
  \bibfield  {author} {\bibinfo {author} {\bibfnamefont {T.~C.}\ \bibnamefont
  {Lubensky}}, \bibinfo {author} {\bibfnamefont {D.}~\bibnamefont {Pettey}},
  \bibinfo {author} {\bibfnamefont {N.}~\bibnamefont {Currier}},\ and\ \bibinfo
  {author} {\bibfnamefont {H.}~\bibnamefont {Stark}},\ }\bibfield  {title}
  {\bibinfo {title} {Topological defects and interactions in nematic
  emulsions},\ }\href {https://doi.org/10.1103/PhysRevE.57.610} {\bibfield
  {journal} {\bibinfo  {journal} {Phys. Rev. E}\ }\textbf {\bibinfo {volume}
  {57}},\ \bibinfo {pages} {610} (\bibinfo {year} {1998})}\BibitemShut
  {NoStop}%
\bibitem [{\citenamefont {Poulin}\ and\ \citenamefont
  {Weitz}(1998)}]{Poulin1998}%
  \BibitemOpen
  \bibfield  {author} {\bibinfo {author} {\bibfnamefont {P.}~\bibnamefont
  {Poulin}}\ and\ \bibinfo {author} {\bibfnamefont {D.~A.}\ \bibnamefont
  {Weitz}},\ }\bibfield  {title} {\bibinfo {title} {\textcolor{black}{Inverted
  and multiple nematic emulsions}},\ }\href
  {https://doi.org/10.1103/PhysRevE.57.626} {\bibfield  {journal} {\bibinfo
  {journal} {Phys. Rev. E}\ }\textbf {\bibinfo {volume} {57}},\ \bibinfo
  {pages} {626} (\bibinfo {year} {1998})}\BibitemShut {NoStop}%
\bibitem [{\citenamefont {Yada}\ \emph {et~al.}(2004)\citenamefont {Yada},
  \citenamefont {Yamamoto},\ and\ \citenamefont {Yokoyama}}]{Yada2004}%
  \BibitemOpen
  \bibfield  {author} {\bibinfo {author} {\bibfnamefont {M.}~\bibnamefont
  {Yada}}, \bibinfo {author} {\bibfnamefont {J.}~\bibnamefont {Yamamoto}},\
  and\ \bibinfo {author} {\bibfnamefont {H.}~\bibnamefont {Yokoyama}},\
  }\bibfield  {title} {\bibinfo {title} {Direct observation of anisotropic
  interparticle forces in nematic colloids with optical tweezers},\ }\href
  {https://doi.org/10.1103/PhysRevLett.92.185501} {\bibfield  {journal}
  {\bibinfo  {journal} {Phys. Rev. Lett.}\ }\textbf {\bibinfo {volume} {92}},\
  \bibinfo {pages} {185501} (\bibinfo {year} {2004})}\BibitemShut {NoStop}%
\bibitem [{\citenamefont {Zapotocky}\ \emph {et~al.}(1999)\citenamefont
  {Zapotocky}, \citenamefont {Ramos}, \citenamefont {Poulin}, \citenamefont
  {Lubensky},\ and\ \citenamefont {Weitz}}]{Zapoitocky1999}%
  \BibitemOpen
  \bibfield  {author} {\bibinfo {author} {\bibfnamefont {M.}~\bibnamefont
  {Zapotocky}}, \bibinfo {author} {\bibfnamefont {L.}~\bibnamefont {Ramos}},
  \bibinfo {author} {\bibfnamefont {P.}~\bibnamefont {Poulin}}, \bibinfo
  {author} {\bibfnamefont {T.~C.}\ \bibnamefont {Lubensky}},\ and\ \bibinfo
  {author} {\bibfnamefont {D.~A.}\ \bibnamefont {Weitz}},\ }\bibfield  {title}
  {\bibinfo {title} {Particle-stabilized defect gel in cholesteric liquid
  crystals},\ }\href {https://doi.org/10.1126/science.283.5399.209} {\bibfield
  {journal} {\bibinfo  {journal} {Science}\ }\textbf {\bibinfo {volume}
  {283}},\ \bibinfo {pages} {209} (\bibinfo {year} {1999})}\BibitemShut
  {NoStop}%
\bibitem [{\citenamefont {Head}\ \emph
  {et~al.}(2024{\natexlab{b}})\citenamefont {Head}, \citenamefont {Negro},
  \citenamefont {Carenza}, \citenamefont {Johnson}, \citenamefont {Keogh},
  \citenamefont {Gonnella}, \citenamefont {Morozov}, \citenamefont {Orlandini},
  \citenamefont {Shendruk}, \citenamefont {Tiribocchi},\ and\ \citenamefont
  {Marenduzzo}}]{Head2024_3}%
  \BibitemOpen
  \bibfield  {author} {\bibinfo {author} {\bibfnamefont {L.~C.}\ \bibnamefont
  {Head}}, \bibinfo {author} {\bibfnamefont {G.}~\bibnamefont {Negro}},
  \bibinfo {author} {\bibfnamefont {L.~N.}\ \bibnamefont {Carenza}}, \bibinfo
  {author} {\bibfnamefont {N.}~\bibnamefont {Johnson}}, \bibinfo {author}
  {\bibfnamefont {R.~R.}\ \bibnamefont {Keogh}}, \bibinfo {author}
  {\bibfnamefont {G.}~\bibnamefont {Gonnella}}, \bibinfo {author}
  {\bibfnamefont {A.}~\bibnamefont {Morozov}}, \bibinfo {author} {\bibfnamefont
  {E.}~\bibnamefont {Orlandini}}, \bibinfo {author} {\bibfnamefont {T.~N.}\
  \bibnamefont {Shendruk}}, \bibinfo {author} {\bibfnamefont {A.}~\bibnamefont
  {Tiribocchi}},\ and\ \bibinfo {author} {\bibfnamefont {D.}~\bibnamefont
  {Marenduzzo}},\ }\bibfield  {title} {\bibinfo {title} {Majorana
  quasiparticles and topological phases in 3d active nematics},\ }\href
  {https://doi.org/10.1073/pnas.2405304121} {\bibfield  {journal} {\bibinfo
  {journal} {Proc. Natl. Acad. Sci. U.S.A.}\ }\textbf {\bibinfo {volume}
  {121}},\ \bibinfo {pages} {e2405304121} (\bibinfo {year}
  {2024}{\natexlab{b}})}\BibitemShut {NoStop}%
\bibitem [{\citenamefont {Carenza}\ \emph
  {et~al.}(2019{\natexlab{a}})\citenamefont {Carenza}, \citenamefont
  {Gonnella}, \citenamefont {Marenduzzo},\ and\ \citenamefont
  {Negro}}]{carenza2019}%
  \BibitemOpen
  \bibfield  {author} {\bibinfo {author} {\bibfnamefont {L.~N.}\ \bibnamefont
  {Carenza}}, \bibinfo {author} {\bibfnamefont {G.}~\bibnamefont {Gonnella}},
  \bibinfo {author} {\bibfnamefont {D.}~\bibnamefont {Marenduzzo}},\ and\
  \bibinfo {author} {\bibfnamefont {G.}~\bibnamefont {Negro}},\ }\bibfield
  {title} {\bibinfo {title} {Rotation and propulsion in {3D} active chiral
  droplets},\ }\href {https://doi.org/10.1073/pnas.1908652116} {\bibfield
  {journal} {\bibinfo  {journal} {Proc. Natl. Acad. Sci. U.S.A.}\ }\textbf
  {\bibinfo {volume} {116}},\ \bibinfo {pages} {22065} (\bibinfo {year}
  {2019}{\natexlab{a}})}\BibitemShut {NoStop}%
\bibitem [{\citenamefont {Ignés-Mullol}\ and\ \citenamefont
  {Sagués}(2020)}]{Mullol2020}%
  \BibitemOpen
  \bibfield  {author} {\bibinfo {author} {\bibfnamefont {J.}~\bibnamefont
  {Ignés-Mullol}}\ and\ \bibinfo {author} {\bibfnamefont {F.}~\bibnamefont
  {Sagués}},\ }\bibfield  {title} {\bibinfo {title} {Active, self-motile, and
  driven emulsions},\ }\href {https://doi.org/10.1016/j.cocis.2020.04.007}
  {\bibfield  {journal} {\bibinfo  {journal} {Curr. Opin. Colloid Interface
  Sci.}\ }\textbf {\bibinfo {volume} {49}},\ \bibinfo {pages} {16} (\bibinfo
  {year} {2020})}\BibitemShut {NoStop}%
\bibitem [{\citenamefont {Guillamat}\ \emph {et~al.}(2018)\citenamefont
  {Guillamat}, \citenamefont {Kos}, \citenamefont {Hardo\"uin}, \citenamefont
  {Ignés-Mullol}, \citenamefont {Ravnik},\ and\ \citenamefont
  {Sagués}}]{Guillamat2018}%
  \BibitemOpen
  \bibfield  {author} {\bibinfo {author} {\bibfnamefont {P.}~\bibnamefont
  {Guillamat}}, \bibinfo {author} {\bibfnamefont {Z.}~\bibnamefont {Kos}},
  \bibinfo {author} {\bibfnamefont {J.}~\bibnamefont {Hardo\"uin}}, \bibinfo
  {author} {\bibfnamefont {J.}~\bibnamefont {Ignés-Mullol}}, \bibinfo {author}
  {\bibfnamefont {M.}~\bibnamefont {Ravnik}},\ and\ \bibinfo {author}
  {\bibfnamefont {F.}~\bibnamefont {Sagués}},\ }\bibfield  {title} {\bibinfo
  {title} {Active nematic emulsions},\ }\href
  {https://doi.org/10.1126/sciadv.aao1470} {\bibfield  {journal} {\bibinfo
  {journal} {Sci. Adv.}\ }\textbf {\bibinfo {volume} {4}},\ \bibinfo {pages}
  {eaao1470} (\bibinfo {year} {2018})}\BibitemShut {NoStop}%
\bibitem [{\citenamefont {Singh}\ \emph {et~al.}(2020)\citenamefont {Singh},
  \citenamefont {Tjhung},\ and\ \citenamefont {Cates}}]{Singh2020}%
  \BibitemOpen
  \bibfield  {author} {\bibinfo {author} {\bibfnamefont {R.}~\bibnamefont
  {Singh}}, \bibinfo {author} {\bibfnamefont {E.}~\bibnamefont {Tjhung}},\ and\
  \bibinfo {author} {\bibfnamefont {M.~E.}\ \bibnamefont {Cates}},\ }\bibfield
  {title} {\bibinfo {title} {Self-propulsion of active droplets without
  liquid-crystalline order},\ }\href
  {https://doi.org/10.1103/PhysRevResearch.2.032024} {\bibfield  {journal}
  {\bibinfo  {journal} {Phys. Rev. Res.}\ }\textbf {\bibinfo {volume} {2}},\
  \bibinfo {pages} {032024} (\bibinfo {year} {2020})}\BibitemShut {NoStop}%
\bibitem [{\citenamefont {Carenza}\ \emph
  {et~al.}(2020{\natexlab{b}})\citenamefont {Carenza}, \citenamefont
  {Gonnella}, \citenamefont {Marenduzzo},\ and\ \citenamefont
  {Negro}}]{carenzaphysicaAchol}%
  \BibitemOpen
  \bibfield  {author} {\bibinfo {author} {\bibfnamefont {L.~N.}\ \bibnamefont
  {Carenza}}, \bibinfo {author} {\bibfnamefont {G.}~\bibnamefont {Gonnella}},
  \bibinfo {author} {\bibfnamefont {D.}~\bibnamefont {Marenduzzo}},\ and\
  \bibinfo {author} {\bibfnamefont {G.}~\bibnamefont {Negro}},\ }\bibfield
  {title} {\bibinfo {title} {Chaotic and periodical dynamics of active chiral
  droplets},\ }\href {https://doi.org/10.1016/j.physa.2020.125025} {\bibfield
  {journal} {\bibinfo  {journal} {Physica A}\ }\textbf {\bibinfo {volume}
  {559}},\ \bibinfo {pages} {125025} (\bibinfo {year}
  {2020}{\natexlab{b}})}\BibitemShut {NoStop}%
\bibitem [{\citenamefont {Tjhung}\ \emph {et~al.}(2017)\citenamefont {Tjhung},
  \citenamefont {Cates},\ and\ \citenamefont {Marenduzzo}}]{tjhung2017}%
  \BibitemOpen
  \bibfield  {author} {\bibinfo {author} {\bibfnamefont {E.}~\bibnamefont
  {Tjhung}}, \bibinfo {author} {\bibfnamefont {M.~E.}\ \bibnamefont {Cates}},\
  and\ \bibinfo {author} {\bibfnamefont {D.}~\bibnamefont {Marenduzzo}},\
  }\bibfield  {title} {\bibinfo {title} {Contractile and chiral activities
  codetermine the helicity of swimming droplet trajectories},\ }\href
  {https://doi.org/10.1073/pnas.1619033114} {\bibfield  {journal} {\bibinfo
  {journal} {Proc. Natl. Acad. Sci. U.S.A.}\ }\textbf {\bibinfo {volume}
  {114}},\ \bibinfo {pages} {4631} (\bibinfo {year} {2017})}\BibitemShut
  {NoStop}%
\bibitem [{\citenamefont {\ifmmode~\check{C}\else \v{C}\fi{}opar}\ \emph
  {et~al.}(2019)\citenamefont {\ifmmode~\check{C}\else \v{C}\fi{}opar},
  \citenamefont {Aplinc}, \citenamefont {Kos}, \citenamefont
  {\ifmmode~\check{Z}\else \v{Z}\fi{}umer},\ and\ \citenamefont
  {Ravnik}}]{Copar2019}%
  \BibitemOpen
  \bibfield  {author} {\bibinfo {author} {\bibfnamefont {S.}~\bibnamefont
  {\ifmmode~\check{C}\else \v{C}\fi{}opar}}, \bibinfo {author} {\bibfnamefont
  {J.}~\bibnamefont {Aplinc}}, \bibinfo {author} {\bibfnamefont {i.~c.~v.}\
  \bibnamefont {Kos}}, \bibinfo {author} {\bibfnamefont {S.}~\bibnamefont
  {\ifmmode~\check{Z}\else \v{Z}\fi{}umer}},\ and\ \bibinfo {author}
  {\bibfnamefont {M.}~\bibnamefont {Ravnik}},\ }\bibfield  {title} {\bibinfo
  {title} {Topology of three-dimensional active nematic turbulence confined to
  droplets},\ }\href {https://doi.org/10.1103/PhysRevX.9.031051} {\bibfield
  {journal} {\bibinfo  {journal} {Phys. Rev. X}\ }\textbf {\bibinfo {volume}
  {9}},\ \bibinfo {pages} {031051} (\bibinfo {year} {2019})}\BibitemShut
  {NoStop}%
\bibitem [{\citenamefont {Nejad}\ and\ \citenamefont
  {Yeomans}(2023)}]{Nejad2023}%
  \BibitemOpen
  \bibfield  {author} {\bibinfo {author} {\bibfnamefont {M.~R.}\ \bibnamefont
  {Nejad}}\ and\ \bibinfo {author} {\bibfnamefont {J.~M.}\ \bibnamefont
  {Yeomans}},\ }\bibfield  {title} {\bibinfo {title} {Spontaneous rotation of
  active droplets in two and three dimensions},\ }\href
  {https://doi.org/10.1103/PRXLife.1.023008} {\bibfield  {journal} {\bibinfo
  {journal} {PRX Life}\ }\textbf {\bibinfo {volume} {1}},\ \bibinfo {pages}
  {023008} (\bibinfo {year} {2023})}\BibitemShut {NoStop}%
\bibitem [{\citenamefont {Negro}\ \emph {et~al.}(2025)\citenamefont {Negro},
  \citenamefont {Head}, \citenamefont {Carenza}, \citenamefont {Shendruk},
  \citenamefont {Marenduzzo}, \citenamefont {Gonnella},\ and\ \citenamefont
  {Tiribocchi}}]{Negro2025}%
  \BibitemOpen
  \bibfield  {author} {\bibinfo {author} {\bibfnamefont {G.}~\bibnamefont
  {Negro}}, \bibinfo {author} {\bibfnamefont {L.~C.}\ \bibnamefont {Head}},
  \bibinfo {author} {\bibfnamefont {L.~N.}\ \bibnamefont {Carenza}}, \bibinfo
  {author} {\bibfnamefont {T.~N.}\ \bibnamefont {Shendruk}}, \bibinfo {author}
  {\bibfnamefont {D.}~\bibnamefont {Marenduzzo}}, \bibinfo {author}
  {\bibfnamefont {G.}~\bibnamefont {Gonnella}},\ and\ \bibinfo {author}
  {\bibfnamefont {A.}~\bibnamefont {Tiribocchi}},\ }\bibfield  {title}
  {\bibinfo {title} {Topology controls flow patterns in active double
  emulsions},\ }\href {https://doi.org/10.1038/s41467-025-56236-8} {\bibfield
  {journal} {\bibinfo  {journal} {Nat. Comm.}\ }\textbf {\bibinfo {volume}
  {16}},\ \bibinfo {pages} {1412} (\bibinfo {year} {2025})}\BibitemShut
  {NoStop}%
\bibitem [{\citenamefont {Alam}\ \emph {et~al.}(2024)\citenamefont {Alam},
  \citenamefont {Najma}, \citenamefont {Singh}, \citenamefont {Laprade},
  \citenamefont {Gajeshwar}, \citenamefont {Yevick}, \citenamefont {Baskaran},
  \citenamefont {Foster},\ and\ \citenamefont {Duclos}}]{Alam2024}%
  \BibitemOpen
  \bibfield  {author} {\bibinfo {author} {\bibfnamefont {S.}~\bibnamefont
  {Alam}}, \bibinfo {author} {\bibfnamefont {B.}~\bibnamefont {Najma}},
  \bibinfo {author} {\bibfnamefont {A.}~\bibnamefont {Singh}}, \bibinfo
  {author} {\bibfnamefont {J.}~\bibnamefont {Laprade}}, \bibinfo {author}
  {\bibfnamefont {G.}~\bibnamefont {Gajeshwar}}, \bibinfo {author}
  {\bibfnamefont {H.~G.}\ \bibnamefont {Yevick}}, \bibinfo {author}
  {\bibfnamefont {A.}~\bibnamefont {Baskaran}}, \bibinfo {author}
  {\bibfnamefont {P.~J.}\ \bibnamefont {Foster}},\ and\ \bibinfo {author}
  {\bibfnamefont {G.}~\bibnamefont {Duclos}},\ }\bibfield  {title} {\bibinfo
  {title} {Active fr\'eedericksz transition in active nematic droplets},\
  }\href {https://doi.org/10.1103/PhysRevX.14.041002} {\bibfield  {journal}
  {\bibinfo  {journal} {Phys. Rev. X}\ }\textbf {\bibinfo {volume} {14}},\
  \bibinfo {pages} {041002} (\bibinfo {year} {2024})}\BibitemShut {NoStop}%
\bibitem [{\citenamefont {Lavrentovich}(2014)}]{Lavrentovich2014}%
  \BibitemOpen
  \bibfield  {author} {\bibinfo {author} {\bibfnamefont {O.~D.}\ \bibnamefont
  {Lavrentovich}},\ }\bibfield  {title} {\bibinfo {title} {Transport of
  particles in liquid crystals},\ }\href {https://doi.org/10.1039/C3SM51628H}
  {\bibfield  {journal} {\bibinfo  {journal} {Soft Matter}\ }\textbf {\bibinfo
  {volume} {10}},\ \bibinfo {pages} {1264} (\bibinfo {year}
  {2014})}\BibitemShut {NoStop}%
\bibitem [{\citenamefont {Tkalec}\ and\ \citenamefont
  {Muševič}(2013)}]{Tkalec2013}%
  \BibitemOpen
  \bibfield  {author} {\bibinfo {author} {\bibfnamefont {U.}~\bibnamefont
  {Tkalec}}\ and\ \bibinfo {author} {\bibfnamefont {I.}~\bibnamefont
  {Muševič}},\ }\bibfield  {title} {\bibinfo {title} {Topology of nematic
  liquid crystal colloids confined to two dimensions},\ }\href
  {https://doi.org/10.1039/C3SM50713K} {\bibfield  {journal} {\bibinfo
  {journal} {Soft Matter}\ }\textbf {\bibinfo {volume} {9}},\ \bibinfo {pages}
  {8140} (\bibinfo {year} {2013})}\BibitemShut {NoStop}%
\bibitem [{\citenamefont {Head}\ \emph
  {et~al.}(2024{\natexlab{c}})\citenamefont {Head}, \citenamefont {Fosado},
  \citenamefont {Marenduzzo},\ and\ \citenamefont {Shendruk}}]{head2024_2}%
  \BibitemOpen
  \bibfield  {author} {\bibinfo {author} {\bibfnamefont {L.~C.}\ \bibnamefont
  {Head}}, \bibinfo {author} {\bibfnamefont {Y.~A.~G.}\ \bibnamefont {Fosado}},
  \bibinfo {author} {\bibfnamefont {D.}~\bibnamefont {Marenduzzo}},\ and\
  \bibinfo {author} {\bibfnamefont {T.~N.}\ \bibnamefont {Shendruk}},\
  }\bibfield  {title} {\bibinfo {title} {Entangled nematic disclinations using
  multi-particle collision dynamics},\ }\href
  {https://doi.org/10.1039/D4SM00436A} {\bibfield  {journal} {\bibinfo
  {journal} {Soft Matter}\ }\textbf {\bibinfo {volume} {20}},\ \bibinfo {pages}
  {7157} (\bibinfo {year} {2024}{\natexlab{c}})}\BibitemShut {NoStop}%
\bibitem [{\citenamefont {{\v C}opar}\ \emph {et~al.}(2015)\citenamefont {{\v
  C}opar}, \citenamefont {Tkalec}, \citenamefont {Mu{\v s}evi{\v c}},\ and\
  \citenamefont {{\v Z}umer}}]{Copar2015}%
  \BibitemOpen
  \bibfield  {author} {\bibinfo {author} {\bibfnamefont {S.}~\bibnamefont {{\v
  C}opar}}, \bibinfo {author} {\bibfnamefont {U.}~\bibnamefont {Tkalec}},
  \bibinfo {author} {\bibfnamefont {I.}~\bibnamefont {Mu{\v s}evi{\v c}}},\
  and\ \bibinfo {author} {\bibfnamefont {S.}~\bibnamefont {{\v Z}umer}},\
  }\bibfield  {title} {\bibinfo {title} {Knot theory realizations in nematic
  colloids},\ }\href {https://doi.org/10.1073/pnas.1417178112} {\bibfield
  {journal} {\bibinfo  {journal} {Proc. Natl. Acad. Sci. U.S.A.}\ }\textbf
  {\bibinfo {volume} {112}},\ \bibinfo {pages} {1675} (\bibinfo {year}
  {2015})}\BibitemShut {NoStop}%
\bibitem [{\citenamefont {Luo}\ \emph {et~al.}(2018)\citenamefont {Luo},
  \citenamefont {Beller}, \citenamefont {Boniello}, \citenamefont {Serra},\
  and\ \citenamefont {Stebe}}]{Luo2018}%
  \BibitemOpen
  \bibfield  {author} {\bibinfo {author} {\bibfnamefont {Y.}~\bibnamefont
  {Luo}}, \bibinfo {author} {\bibfnamefont {D.~A.}\ \bibnamefont {Beller}},
  \bibinfo {author} {\bibfnamefont {G.}~\bibnamefont {Boniello}}, \bibinfo
  {author} {\bibfnamefont {F.}~\bibnamefont {Serra}},\ and\ \bibinfo {author}
  {\bibfnamefont {K.~J.}\ \bibnamefont {Stebe}},\ }\bibfield  {title} {\bibinfo
  {title} {Tunable colloid trajectories in nematic liquid crystals near wavy
  walls},\ }\href {https://doi.org/10.1038/s41467-018-06054-y} {\bibfield
  {journal} {\bibinfo  {journal} {Nat. Commun.}\ }\textbf {\bibinfo {volume}
  {9}},\ \bibinfo {pages} {3841} (\bibinfo {year} {2018})}\BibitemShut
  {NoStop}%
\bibitem [{\citenamefont {Wamsler}\ \emph {et~al.}(2024)\citenamefont
  {Wamsler}, \citenamefont {Head},\ and\ \citenamefont
  {Shendruk}}]{Wamsler2024}%
  \BibitemOpen
  \bibfield  {author} {\bibinfo {author} {\bibfnamefont {K.}~\bibnamefont
  {Wamsler}}, \bibinfo {author} {\bibfnamefont {L.~C.}\ \bibnamefont {Head}},\
  and\ \bibinfo {author} {\bibfnamefont {T.~N.}\ \bibnamefont {Shendruk}},\
  }\bibfield  {title} {\bibinfo {title} {Lock-key microfluidics: simulating
  nematic colloid advection along wavy-walled channels},\ }\href
  {https://doi.org/10.1039/D3SM01536J} {\bibfield  {journal} {\bibinfo
  {journal} {Soft Matter}\ }\textbf {\bibinfo {volume} {20}},\ \bibinfo {pages}
  {3954} (\bibinfo {year} {2024})}\BibitemShut {NoStop}%
\bibitem [{\citenamefont {Chandler}\ and\ \citenamefont
  {Spagnolie}(2024{\natexlab{b}})}]{Chandler2024}%
  \BibitemOpen
  \bibfield  {author} {\bibinfo {author} {\bibfnamefont {T.~G.~J.}\
  \bibnamefont {Chandler}}\ and\ \bibinfo {author} {\bibfnamefont {S.~E.}\
  \bibnamefont {Spagnolie}},\ }\bibfield  {title} {\bibinfo {title} {Rigid and
  deformable bodies in nematic liquid crystals},\ }\href
  {https://doi.org/10.1103/PhysRevFluids.9.110511} {\bibfield  {journal}
  {\bibinfo  {journal} {Phys. Rev. Fluids}\ }\textbf {\bibinfo {volume} {9}},\
  \bibinfo {pages} {110511} (\bibinfo {year} {2024}{\natexlab{b}})}\BibitemShut
  {NoStop}%
\bibitem [{\citenamefont {Singh}\ and\ \citenamefont
  {Chaudhuri}(2024)}]{singh2024}%
  \BibitemOpen
  \bibfield  {author} {\bibinfo {author} {\bibfnamefont {C.}~\bibnamefont
  {Singh}}\ and\ \bibinfo {author} {\bibfnamefont {A.}~\bibnamefont
  {Chaudhuri}},\ }\bibfield  {title} {\bibinfo {title} {Anomalous dynamics of a
  passive droplet in active turbulence},\ }\href
  {https://doi.org/https://doi.org/10.1038/s41467-024-47727-1} {\bibfield
  {journal} {\bibinfo  {journal} {Nat. Commun.}\ }\textbf {\bibinfo {volume}
  {15}},\ \bibinfo {pages} {3704} (\bibinfo {year} {2024})}\BibitemShut
  {NoStop}%
\bibitem [{\citenamefont {Foffano}\ \emph {et~al.}(2019)\citenamefont
  {Foffano}, \citenamefont {Lintuvuori}, \citenamefont {Stratford},
  \citenamefont {Cates},\ and\ \citenamefont {Marenduzzo}}]{foffano2019}%
  \BibitemOpen
  \bibfield  {author} {\bibinfo {author} {\bibfnamefont {G.}~\bibnamefont
  {Foffano}}, \bibinfo {author} {\bibfnamefont {J.~S.}\ \bibnamefont
  {Lintuvuori}}, \bibinfo {author} {\bibfnamefont {K.}~\bibnamefont
  {Stratford}}, \bibinfo {author} {\bibfnamefont {M.~E.}\ \bibnamefont
  {Cates}},\ and\ \bibinfo {author} {\bibfnamefont {D.}~\bibnamefont
  {Marenduzzo}},\ }\bibfield  {title} {\bibinfo {title} {Dynamic clustering and
  re-dispersion in concentrated colloid-active gel composites},\ }\href
  {https://doi.org/10.1039/C9SM01249D} {\bibfield  {journal} {\bibinfo
  {journal} {Soft Matter}\ }\textbf {\bibinfo {volume} {15}},\ \bibinfo {pages}
  {6896} (\bibinfo {year} {2019})}\BibitemShut {NoStop}%
\bibitem [{\citenamefont {Fily}\ and\ \citenamefont
  {Marchetti}(2012)}]{fily2012}%
  \BibitemOpen
  \bibfield  {author} {\bibinfo {author} {\bibfnamefont {Y.}~\bibnamefont
  {Fily}}\ and\ \bibinfo {author} {\bibfnamefont {M.~C.}\ \bibnamefont
  {Marchetti}},\ }\bibfield  {title} {\bibinfo {title} {Athermal {P}hase
  {S}eparation of {S}elf-{P}ropelled {P}articles with {N}o {A}lignment},\
  }\href {https://doi.org/10.1103/PhysRevLett.108.235702} {\bibfield  {journal}
  {\bibinfo  {journal} {Phys. Rev. Lett.}\ }\textbf {\bibinfo {volume} {108}},\
  \bibinfo {pages} {235702} (\bibinfo {year} {2012})}\BibitemShut {NoStop}%
\bibitem [{\citenamefont {Gonnella}\ \emph {et~al.}(2015)\citenamefont
  {Gonnella}, \citenamefont {Marenduzzo}, \citenamefont {Suma},\ and\
  \citenamefont {Tiribocchi}}]{gonnellaMIPS}%
  \BibitemOpen
  \bibfield  {author} {\bibinfo {author} {\bibfnamefont {G.}~\bibnamefont
  {Gonnella}}, \bibinfo {author} {\bibfnamefont {D.}~\bibnamefont
  {Marenduzzo}}, \bibinfo {author} {\bibfnamefont {A.}~\bibnamefont {Suma}},\
  and\ \bibinfo {author} {\bibfnamefont {A.}~\bibnamefont {Tiribocchi}},\
  }\bibfield  {title} {\bibinfo {title} {Motility-induced phase separation and
  coarsening in active matter},\ }\href
  {https://doi.org/10.1016/j.crhy.2015.03.004} {\bibfield  {journal} {\bibinfo
  {journal} {C. R. Phys.}\ }\textbf {\bibinfo {volume} {16}},\ \bibinfo {pages}
  {316} (\bibinfo {year} {2015})}\BibitemShut {NoStop}%
\bibitem [{\citenamefont {Semeraro}\ \emph {et~al.}(2024)\citenamefont
  {Semeraro}, \citenamefont {Negro}, \citenamefont {Suma}, \citenamefont
  {Corberi},\ and\ \citenamefont {Gonnella}}]{Semeraro_2024}%
  \BibitemOpen
  \bibfield  {author} {\bibinfo {author} {\bibfnamefont {M.}~\bibnamefont
  {Semeraro}}, \bibinfo {author} {\bibfnamefont {G.}~\bibnamefont {Negro}},
  \bibinfo {author} {\bibfnamefont {A.}~\bibnamefont {Suma}}, \bibinfo {author}
  {\bibfnamefont {F.}~\bibnamefont {Corberi}},\ and\ \bibinfo {author}
  {\bibfnamefont {G.}~\bibnamefont {Gonnella}},\ }\bibfield  {title} {\bibinfo
  {title} {Entropy production of active brownian particles going from liquid to
  hexatic and solid phases},\ }\href {https://doi.org/10.1209/0295-5075/ad895e}
  {\bibfield  {journal} {\bibinfo  {journal} {EPL}\ }\textbf {\bibinfo {volume}
  {148}},\ \bibinfo {pages} {37001} (\bibinfo {year} {2024})}\BibitemShut
  {NoStop}%
\bibitem [{\citenamefont {Redner}\ \emph {et~al.}(2013)\citenamefont {Redner},
  \citenamefont {Hagan},\ and\ \citenamefont {A.Baskaran}}]{rednerMIPS}%
  \BibitemOpen
  \bibfield  {author} {\bibinfo {author} {\bibfnamefont {G.}~\bibnamefont
  {Redner}}, \bibinfo {author} {\bibfnamefont {M.}~\bibnamefont {Hagan}},\ and\
  \bibinfo {author} {\bibnamefont {A.Baskaran}},\ }\bibfield  {title} {\bibinfo
  {title} {Structure and dynamics of a phase-separating active colloidal
  fluid},\ }\href {https://doi.org/10.1103/PhysRevLett.110.055701} {\bibfield
  {journal} {\bibinfo  {journal} {Phys. Rev. Lett.}\ }\textbf {\bibinfo
  {volume} {110}},\ \bibinfo {pages} {055701} (\bibinfo {year}
  {2013})}\BibitemShut {NoStop}%
\bibitem [{\citenamefont {Buttinoni}\ \emph {et~al.}(2013)\citenamefont
  {Buttinoni}, \citenamefont {Bialk\'e}, \citenamefont {K\"ummel},
  \citenamefont {L\"owen}, \citenamefont {Bechinger},\ and\ \citenamefont
  {Speck}}]{buttSPP}%
  \BibitemOpen
  \bibfield  {author} {\bibinfo {author} {\bibfnamefont {I.}~\bibnamefont
  {Buttinoni}}, \bibinfo {author} {\bibfnamefont {J.}~\bibnamefont {Bialk\'e}},
  \bibinfo {author} {\bibfnamefont {F.}~\bibnamefont {K\"ummel}}, \bibinfo
  {author} {\bibfnamefont {H.}~\bibnamefont {L\"owen}}, \bibinfo {author}
  {\bibfnamefont {C.}~\bibnamefont {Bechinger}},\ and\ \bibinfo {author}
  {\bibfnamefont {T.}~\bibnamefont {Speck}},\ }\bibfield  {title} {\bibinfo
  {title} {Dynamical clustering and phase separation in suspensions of
  self-propelled colloidal particles},\ }\href
  {https://doi.org/10.1103/PhysRevLett.110.238301} {\bibfield  {journal}
  {\bibinfo  {journal} {Phys. Rev. Lett.}\ }\textbf {\bibinfo {volume} {110}},\
  \bibinfo {pages} {238301} (\bibinfo {year} {2013})}\BibitemShut {NoStop}%
\bibitem [{\citenamefont {Cates}\ and\ \citenamefont
  {Tailleur}(2015)}]{cates2015}%
  \BibitemOpen
  \bibfield  {author} {\bibinfo {author} {\bibfnamefont {M.~E.}\ \bibnamefont
  {Cates}}\ and\ \bibinfo {author} {\bibfnamefont {J.}~\bibnamefont
  {Tailleur}},\ }\bibfield  {title} {\bibinfo {title} {Motility-induced phase
  separation},\ }\href
  {https://doi.org/10.1146/annurev-conmatphys-031214-014710} {\bibfield
  {journal} {\bibinfo  {journal} {Annu. Rev. Condens. Matter Phys.}\ }\textbf
  {\bibinfo {volume} {6}},\ \bibinfo {pages} {219} (\bibinfo {year}
  {2015})}\BibitemShut {NoStop}%
\bibitem [{\citenamefont {de~Gennes}\ and\ \citenamefont
  {Prost}(1993)}]{degennes}%
  \BibitemOpen
  \bibfield  {author} {\bibinfo {author} {\bibfnamefont {P.~G.}\ \bibnamefont
  {de~Gennes}}\ and\ \bibinfo {author} {\bibfnamefont {J.}~\bibnamefont
  {Prost}},\ }\href@noop {} {\emph {\bibinfo {title} {The Physics of Liquid
  Crystals}}}\ (\bibinfo  {publisher} {Clarendon Press, Oxford, 2nd edn},\
  \bibinfo {year} {1993})\BibitemShut {NoStop}%
\bibitem [{\citenamefont {Succi}(2018)}]{succi2018}%
  \BibitemOpen
  \bibfield  {author} {\bibinfo {author} {\bibfnamefont {S.}~\bibnamefont
  {Succi}},\ }\href@noop {} {\emph {\bibinfo {title} {The lattice Boltzmann
  equation: for complex states of flowing matter}}}\ (\bibinfo  {publisher}
  {Oxford University Press, Oxford},\ \bibinfo {year} {2018})\BibitemShut
  {NoStop}%
\bibitem [{\citenamefont {Negro}\ \emph {et~al.}(2023)\citenamefont {Negro},
  \citenamefont {Carenza}, \citenamefont {Gonnella}, \citenamefont {Mackay},
  \citenamefont {Morozov},\ and\ \citenamefont {Marenduzzo}}]{Negro2023}%
  \BibitemOpen
  \bibfield  {author} {\bibinfo {author} {\bibfnamefont {G.}~\bibnamefont
  {Negro}}, \bibinfo {author} {\bibfnamefont {L.~N.}\ \bibnamefont {Carenza}},
  \bibinfo {author} {\bibfnamefont {G.}~\bibnamefont {Gonnella}}, \bibinfo
  {author} {\bibfnamefont {F.}~\bibnamefont {Mackay}}, \bibinfo {author}
  {\bibfnamefont {A.}~\bibnamefont {Morozov}},\ and\ \bibinfo {author}
  {\bibfnamefont {D.}~\bibnamefont {Marenduzzo}},\ }\bibfield  {title}
  {\bibinfo {title} {Yield-stress transition in suspensions of deformable
  droplets},\ }\href {https://doi.org/10.1126/sciadv.adf8106} {\bibfield
  {journal} {\bibinfo  {journal} {Sci. Adv.}\ }\textbf {\bibinfo {volume}
  {9}},\ \bibinfo {pages} {eadf8106} (\bibinfo {year} {2023})}\BibitemShut
  {NoStop}%
\bibitem [{\citenamefont {Carenza}\ \emph
  {et~al.}(2019{\natexlab{b}})\citenamefont {Carenza}, \citenamefont
  {Gonnella}, \citenamefont {Lamura}, \citenamefont {Negro},\ and\
  \citenamefont {Tiribocchi}}]{Carenza2019_2}%
  \BibitemOpen
  \bibfield  {author} {\bibinfo {author} {\bibfnamefont {L.~N.}\ \bibnamefont
  {Carenza}}, \bibinfo {author} {\bibfnamefont {G.}~\bibnamefont {Gonnella}},
  \bibinfo {author} {\bibfnamefont {A.}~\bibnamefont {Lamura}}, \bibinfo
  {author} {\bibfnamefont {G.}~\bibnamefont {Negro}},\ and\ \bibinfo {author}
  {\bibfnamefont {A.}~\bibnamefont {Tiribocchi}},\ }\bibfield  {title}
  {\bibinfo {title} {Lattice boltzmann methods and active fluids},\ }\href
  {https://doi.org/10.1140/epje/i2019-11843-6} {\bibfield  {journal} {\bibinfo
  {journal} {Eur. Phys. J. E}\ }\textbf {\bibinfo {volume} {42}},\ \bibinfo
  {pages} {81} (\bibinfo {year} {2019}{\natexlab{b}})}\BibitemShut {NoStop}%
\end{thebibliography}%

\end{document}



\title{Supplementary Information for: Activity drives self-assembly of passive soft inclusions in active nematics}

\author{Ahmet Umut Akduman}
\thanks{These authors contributed equally.}
 \affiliation{Physics Department, College of Sciences, Ko{\c c} University, Rumelifeneri Yolu 34450 Sar\i{}yer, Istanbul,  T{\" u}rkiye}
 
\author{Yusuf Sar\i{}yar}
\thanks{These authors contributed equally.}
 \affiliation{Physics Department, College of Sciences, Ko{\c c} University, Rumelifeneri Yolu 34450 Sar\i{}yer, Istanbul,  T{\" u}rkiye}

\author{Giuseppe Negro}
\email{giuseppe.negro@ed.ac.uk}
 \affiliation{SUPA, School of Physics and Astronomy, University of Edinburgh, Peter Guthrie Tait Road, Edinburgh, EH9 3FD, UK}
 
\author{Livio Nicola Carenza}
\email{lcarenza@ku.edu.tr}
 \affiliation{Physics Department, College of Sciences, Ko{\c c} University, Rumelifeneri Yolu 34450 Sar\i{}yer, Istanbul,  T{\" u}rkiye}


\maketitle

\subsection*{Movies description}
\begin{itemize}
    \item \textbf{\href{https://youtube.com/shorts/tsXMZsQl3Ts?feature=share}{Movie~S1}. Quiescent State.} The movie shows the time evolution of the total concentration field $\phi$ for a system of $N=20$ droplets ($\Phi=0.09$) at activity $\zeta=10^{-4}$. The purple and yellow regions correspond to $\phi=0$ and $\phi=2$, respectively. The director field is shown with white dashes. During the evolution of the system, the droplets remain predominantly stationary, undergoing only small rearrangements.
    
    \item \textbf{\href{https://youtube.com/shorts/mcLeJ4XFR68?feature=share}{Movie~S2}. Clustering State.} The movie shows the time evolution of the total concentration field $\phi$ for a system of $N=25$ droplets ($\Phi=0.11$) at activity $\zeta=4\times 10^{-3}$. The purple and yellow regions correspond to $\phi=0$ and $\phi=2$, respectively. The director field is shown with white rods. Droplets form dynamical clusters. However, because of disruption of the active jets, cluster growth is arrested, with clusters continuously assembling and disassembling over time.

    \item \textbf{\href{https://youtube.com/shorts/DL9Mmr0DLfs?feature=share}{Movie~S3}. Gelation State.} The movie shows the time evolution of the total concentration field $\phi$ for a system of $N=50$ droplets ($\Phi=0.22$) at activity $\zeta=2\times 10^{-3}$. The purple and yellow regions correspond to $\phi=0$ and $\phi=2$, respectively. The director field is shown with white rods. Initially, the droplets move in a dynamic clustering behavior and exhibit significant translational motion. As time progresses, these clusters merge and grow larger, gradually forming a continuous percolating network. At long times, the network stabilizes and the droplets reach a stationary configuration. This configuration resembles a gel-like structure, with only small rearrangements occurring within the percolating network.
    
    \item \textbf{\href{https://youtube.com/shorts/mi0Bww6evIU?feature=share}{Movie~S4}. Active Turbulence State.} The movie shows the time evolution of the total concentration field $\phi$ for a system of $N=20$ droplets ($\Phi=0.09$) at activity $\zeta=4\times 10^{-2}$. The purple and yellow regions correspond to $\phi=0$ and $\phi=2$, respectively. As the active flow becomes too large, it prevents the formation of clusters, with only transient dimers occasionally forming. The droplet dynamics are highly chaotic, leading to the absence of any persistent or coherent structures.
    
    \item \textbf{\href{https://youtube.com/shorts/e-iF8TSjn9U?feature=share}{Movie~S5}. Inverse MIPS State.} The movie shows the time evolution of the total concentration field $\phi$ for a system of $N=40$ droplets ($\Phi=0.18$) at activity $\zeta=7\times 10^{-2}$. The purple and yellow regions correspond to $\phi=0$ and $\phi=3.7$, respectively. The system begins with the formation of small, transient clusters while occasionally forming chain-like structures. Upon collisions and interactions mediated by the elasticity of liquid crystal, droplets' mobility decreases which leads to local accumulation, and eventually a fully separated clustered state.

    \item \textbf{\href{https://youtube.com/shorts/SGwrGd3eXkY?feature=share}{Movie~S6}. Abrupt Activity Reduction: Clustered State.} The movie shows the time evolution of a system of $N=40$ droplets ($\Phi=0.18$). The purple and yellow regions correspond to $\phi=0$ and $\phi=3.7$, respectively. The Initial configuration is obtained by increasing activity until $\zeta = 7\times 10^{-2}$. Starting from this configuration we then proceed to reduce the activity at rate $-\frac{d\zeta}{dt}=15\times 10^{-8}$. At this fast decay rate of activity, the system maintains its clustered configuration through persistent defect-mediated bonds, and a hexatic matrix of defects between the droplets is formed.
    
    \item \textbf{\href{https://youtube.com/shorts/NsRfsqugN5E?feature=share}{Movie~S7}. Intermediate Activity Reduction: Partially Clustered State.} The movie shows the time evolution of a system of $N=40$ droplets ($\Phi=0.18$). The purple and yellow regions correspond to $\phi=0$ and $\phi=3.7$, respectively. The initial configuration is obtained by increasing activity until $\zeta = 7\times 10^{-2}$. Starting from this configuration we then proceed to reduce the activity at rate $-\frac{d\zeta}{dt}=2\times 10^{-8}$. At this intermediate decay rate of activity, the fully separated cluster starts to melt as peripheral droplets becomes vulnerable to the advective action of turbulent flows. As a result, the initial cluster persists in coexistence with a dispersed suspension of droplets.

    \item \textbf{\href{https://youtube.com/shorts/ruMld3cctlc?feature=share}{Movie~S8}. Hysteresis: Melted State.} The movie shows the time evolution of a system of $N=40$ droplets ($\Phi=0.18$).  The purple and yellow regions correspond to $\phi=0$ and $\phi=3.7$, respectively. The initial configuration is obtained by increasing activity until $\zeta = 7\times 10^{-2}$. Starting from this separated configuration we then proceed to reduce the activity at rate $-\frac{d\zeta}{dt}=10^{-8}$. At this slow decay rate of activity, the system adiabatically returns to its completely disordered configuration through complete melting.
\end{itemize}

